\documentstyle[epsf]{mn}

\begin{document}
\newcommand{\be}{\begin{equation}}
\newcommand{\ee}{\end{equation}}
\newcommand{\ba}{\begin{eqnarray}}
\newcommand{\ea}{\end{eqnarray}}
\def\mpc{h^{-1} {\rm{Mpc}}}
\def\kpc{h^{-1} {\rm{kpc}}}
\def\up{h^{-3} {\rm{Mpc^3}}}
\def\uk{h {\rm{Mpc^{-1}}}}
\def\lsim{\mathrel{\hbox{\rlap{\hbox{\lower4pt\hbox{$\sim$}}}\hbox{$<$}}}}
\def\gsim{\mathrel{\hbox{\rlap{\hbox{\lower4pt\hbox{$\sim$}}}\hbox{$>$}}}}
\def\kms {\rm{km~s^{-1}}}
\def\masa{h^{-1}{{\cal M}_{\odot}}}\def\apj {ApJ}
\def\aj {AJ}
\def\aa {A \& A}
\def\mnras {MNRAS}
\newcommand{\mincir}{\raise
-2.truept\hbox{\rlap{\hbox{$\sim$}}\raise5.truept\hbox{$<$}\ }}
\newcommand{\magcir}{\raise
-2.truept\hbox{\rlap{\hbox{$\sim$}}\raise5.truept\hbox{$>$}\ }}
\newcommand{\minmag}{\raise
-2.truept\hbox{\rlap{\hbox{$<$}}\raise6.truept\hbox{$<$}\ }}

\title[Dark Energy with HII galaxies \& X-ray AGN]
{A Strategy to Measure the Dark Energy Equation of State using
  the HII galaxy Hubble Function \& X-ray AGN Clustering: 
Preliminary Results.}
%
%

\author[Plionis, M. et al.]
{
  \parbox[t]{\textwidth}
  {
  M. Plionis${^{1,2}}$, R. Terlevich$^{2}$, S. Basilakos$^3$,
  F. Bresolin$^4$, E. Terlevich$^2$, J. Melnick$^5$, 
  R. Chavez$^2$}
  \vspace*{6pt}\\ 
  \parbox[t]{15 cm}
  {
    $^1$ Institute of Astronomy \& Astrophysics, National Observatory of Athens,
    Palaia Penteli 152 36, Athens, Greece.\\
    $^2$ Instituto Nacional de Astrof\'{\i}sica Optica y Electr\'onica, AP 51
    y 216, 72000, Puebla, M\'exico.\\
    $^3$Academy of Athens, Research Center for Astronomy and Applied Mathematics,
 Soranou Efesiou 4, 11527, Athens, Greece \\
    $^4$ Institute for Astronomy of the University of Hawaii, 2680
    Woodlawn Drive, 96822 Honolulu, HI USA \\
    $^5$ European Southern Observatory, Alonso de Cordoba 3107, Santiago, Chile
  }
}
\date{\today}

\maketitle

\begin{abstract}
We explore the possibility of setting stringent constraints to the Dark Energy equation of state using
alternative cosmic tracers like: (a) the Hubble relation using
HII galaxies, which can be observed at much higher redshifts ($ z
\mincir 3.5$) than those currently traced by SNIa samples, and (b) the
large-scale structure using the clustering of X-ray selected AGN,
which have a redshift distribution peaking at $z\sim 1$.

In this paper we use 
extensive Monte-Carlo simulations to define the optimal strategy for the recovery of the dark-energy 
equation of state using the high redshift ($z\magcir 2$) Hubble
relation, but accounting also for the effects of gravitational lensing,
which for such high redshifts can significantly affect
the derived cosmological constraints.
We investigate the size of the sample of high-$z$
HII-galaxies needed to provide  useful constraints in the Dark
Energy equation of state. Based on a ``Figure of Merit'' analysis, 
we provide estimates for the number of  $2\mincir z\mincir 3.5$
tracers needed to reduce the cosmological solution space,
presently provided by the {\em Constitution} SNIa set, by a desired factor. 
The analysis is given for any level of rms distance
modulus uncertainty and we find that an expected reduction (i.e. by $\sim
20\%-40\%$) of the current level of HII-galaxy based distance modulus uncertainty 
does not provide a significant improvement in the derived 
cosmological constraints. It is much more efficient to increase the number of tracers 
than to reduce their individual uncertainties.

Finally, we propose a framework to put  constraints on the dark energy equation
of state by using the joint likelihood of the X-ray AGN
clustering and of the Hubble relation cosmological analyses.
A preliminary joint analysis using the
X-ray AGN clustering of the 2XMM survey and the Hubble relation
of the {\em Constitution} SNIa set provide:
$\Omega_{\rm m}= 0.31\pm 0.01$ and w$=-1.06\pm 0.05$. We also
find that the joint SNIa-2XMM analysis provides significantly more
stringent cosmological constraints,
increasing the Figure of Merit by a factor $\sim 2$, 
with respect to that of the joint SNIa-BAO analysis.
\end{abstract}

\section{Introduction}
We live in a very exciting period for our understanding of the
Cosmos. Over the past decade the accumulation and detailed analyses of
high quality cosmological data (eg., supernovae type Ia, CMB
temperature fluctuations, galaxy clustering, high $z$ clusters of
galaxies, etc.) have strongly suggested that we live in a 
flat and accelerating universe, which contains at least some sort of
cold dark matter to explain the clustering of extragalactic sources,
and an extra component which acts as having a negative pressure, as for
example the energy of the vacuum (or in a more general setting the so
called {\em dark energy}), to explain the observed accelerated cosmic
expansion  (eg. Riess, et al. 1998; 2004; 2007, Perlmutter et
al. 1999; Spergel et al. 2003, 2007, Tonry et al. 2003; Schuecker et
al. 2003; Tegmark et al. 2004; Seljak et al. 2004; Allen et al. 2004;
Basilakos \& Plionis 2005; 2006; 2009; Blake et al. 2007; Wood-Vasey et
al. 2007, Davis et al. 2007; Kowalski et al. 2008, Komatsu et
al. 2009; Hicken et al. 2009, Amanullah et al. 2010, etc).

Due to the absence of a well-motivated fundamental theory, there have
been many theoretical speculations regarding the nature of the 
{\em dark energy} (hereafter DE), on whether it is a cosmological
constant or a field that provides a time varying equation of
state, usually parametrized by: 
\begin{equation}
p_Q= {\rm w}(z) \rho_Q\;,
\end{equation} 
with $p_Q$ and $\rho_Q$ the pressure and density of the exotic dark
energy fluid. For a large class of DE models, we have:
\be 
{\rm w}(z)={\rm w}_0 + {\rm w}_1 f(z) \;,
\label{eqstatez}
\ee 
with w$_0=$w$(z=0)$ and $f(z)$ an increasing function of redshift. A
particular example of w$(z)$ is its 1st order Taylor's expansion
around w(0), which provides $f(z)=z/(1+z)$, ie., the so-called CPL form 
of the DE equation of state (Chevalier \& Polarski 2001, Linder 2003; see also 
Peebles \& Ratra 2003, Dicus \& Repko 2004; Wang \& Mukherjee 2006). Of
course, it could also be conceived that 
the equation of state parameter does not evolve
cosmologically ({\em quintessence} Dark Energy model; QDE). 

It is clear that one of the most important questions in
Cosmology and cosmic structure formation is related to the nature of
{\em dark energy} (as well as whether it is the sole explanation of the
observed accelerated expansion of the Universe) and its interpretation
within a fundamental physical theory (eg., Albrecht et al. 2006; Peacock
et al. 2006). To this end, a large number of
very expensive experiments are proposed and/or are at various stages of
development, viz the {\em Dark Energy Survey}: 
{\tt http://www.darkenergysurvey.org/}, the {\em Joint Dark Energy
  Mission}: {\tt http://jdem.gsfc.nasa.gov/}, {\em HETDEX} 
{\tt http://www.as.utexas.edu/hetdex/}, 
{\em Pan-STARRS}: {\tt http://pan-starrs.ifa.hawaii.edu}, {\em Euclid}: {\tt
  http://sci.esa.int/euclid/}, {\em Wfirst}: {\tt
  http://wfirst.gsfc.nasa.gov}, etc.

Therefore, the paramount importance of the detection and
quantification of DE for our understanding of the cosmos and
for fundamental theories implies that the results of the different
experiments should not only be scrutinized, but alternative, even
higher-risk, methods to measure DE should be developed and
applied as well.
 It is within this paradigm that our current work falls.
Indeed, we wish to constrain the DE equation of state
using, individually and in 
combination, the Hubble relation and large-scale structure (clustering) methods, but
utilizing alternative cosmic tracers for both of these components.
 
Thus, we will trace the Hubble relation using HII 
galaxies, which can be observed at higher redshifts than reached
by current SNIa surveys to distances where the Hubble relation
is more sensitive to the cosmological parameters. The HII galaxies can
be used as standard candles (Melnick, Terlevich \& Terlevich 2000,
Melnick 2003; Siegel et al. 2005; Plionis et al. 2009) 
due to the correlation between their velocity
dispersion, $H_{\beta}$ luminosity and metallicity (Melnick 1978, Terlevich \&
Melnick 1981, Melnick, Terlevich \& Moles 1988). Furthermore, the use
of such alternative high $z$ tracer will enable us to check the SNIa results
and lift any doubts that arise from the fact that they are the only
tracers of the Hubble relation used to-date (for possible usage of GRBs
see for example, Ghirlanda et al. 2006, 2009; Basilakos \& Perivolaropoulos
2008 and references therein)
\footnote{
GRBs appear to be anything but standard candles, having a very wide
range of isotropic equivalent luminosities and energy outputs.
Nevertheless, correlations between various properties of
the prompt emission and in some cases also the afterglow emission
have been used to determine their distances.
A serious problem that hampers a straightforward use of GRBs as
Cosmological probes is the intrinsic
faintness of the nearby events, a fact which introduces a bias towards low (or high)
values of GRB observables and therefore the extrapolation
of their correlations to low-$z$ events is faced with
serious problems.
One might also expect a significant evolution of the
intrinsic properties of GRBs with redshift (also between
intermediate and high redshifts) which can be hard to disentangle
from cosmological effects. Finally, even if a reliable scaling relation
can be identified and used, the scatter in the resulting
luminosity and thus distance modulus is still fairly large.}.

Additionally, we use X-ray selected AGN at a median
redshift of $\sim 1$, which is roughly the
peak of their redshift distribution (see Basilakos et al. 2004; 2005,
Miyaji et al. 2007), in order to determine their clustering pattern
and compare it with that predicted by different cosmological models
(see Matsubara 2004). 

Although each of the previously discussed components of our project
(Hubble relation using HII galaxies and angular/spatial
clustering of X-ray AGN) will provide interesting and relatively
stringent constraints on the cosmological parameters, especially under
our anticipation that we will reduce significantly the corresponding
random and systematic errors, it is the combined likelihood of these
two type of analyses that enables us to break the known
degeneracies between cosmological parameters and determine with great
accuracy the DE equation of state (see 
Basilakos \& Plionis 2005; 2006; 2009).

Below we present the basic methodology and expectations of the 
two components of our method. In section 2 we
present the details of the first component where we develop a
Monte-Carlo simulation approach designed to ultimately provide a rule
of thumb of how many HII galaxies we need to obtain a
particular level of the DE equation of state parameter uncertainty. We also develop a
method to account for the effects of gravitational lensing, which at
such high redshifts are significant.
In section 3 we present the details of the second component 
and in section 4 we present an example of joining the two
components to provide cosmological constraints. The conclusions are
listed in section 5. 

\section{Cosmological Parameters from the Hubble Relation}
In the matter dominated epoch and in flat
universes, the Hubble relation depends on the cosmological
parameters via the following equation:
\be
H(z) = H_{0} E(z)
\ee
with
\be\label{eq:HR}
E^2(z)=\Omega_m (1+z)^3 + \Omega_k (1+z)^2 + \Omega_Q \exp \left[3 \int_0^z
  \frac{1+{\rm w}(x)}{1+x} {\rm d}x\right] \;,
\ee
which is simply derived from Friedman's equation. We remind the reader
that $\Omega_m$, $\Omega_k$ and $\Omega_Q (\equiv 1-\Omega_{m}-\Omega_k)$ are 
the present fractional contributions to the
total cosmic mass-energy density of the matter, the spatial curvature and dark energy source
terms, respectively.
\begin{figure*}
\centering
\mbox{\epsfxsize=14cm \epsffile{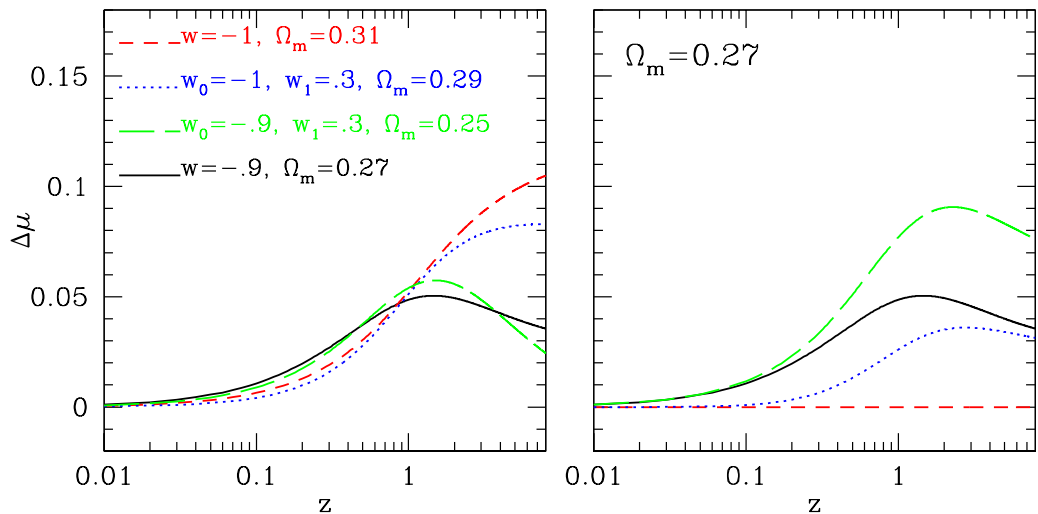}}
\caption{{\em Left Panel:} The expected distance modulus difference between the
  DE models shown and the reference
  $\Lambda$-model (w$=-1$) with $\Omega_{m}=0.27$. {\em Right Panel:}
  The expected distance modulus differences once the $\Omega_m$-w$(z)$
  degeneracy is broken (imposing the same $\Omega_m$ value as in the comparison model).}
\end{figure*} 

Supernovae SNIa are considered standard candles at peak luminosity
and therefore they have been used not only to determine the Hubble constant (at
relatively low redshifts) but also to trace the curvature of the
Hubble relation at high redshifts (see Riess et al. 1998, 2004, 2007; Perlmutter et
al. 1998, 1999; Tonry et al. 2003; Astier et al. 2006; Wood-Vasey et
al. 2007; Davis et al. 2007; Kowalski et al 2008; Hicken et al. 2009; 
Amanullah et al. 2010; Wang, Li \& Li 2011; Kim 2011; March et al. 2011; Adak,
Bandopadhyay, Majumdar 2011).
In practice one relates the distance modulus of the SNIa to its luminosity
distance, $d_L$, through which the cosmological parameters enter:
\be\label{eq:dm}
\mu = m-M = 5 \log d_L + 25 
\ee 
and
\be
d_{L} =  \frac{c(1+z)}{H_0 \sqrt{\Omega_{k}}}
\sinh\left[\sqrt{\Omega_{k}} \int_0^z \frac{{\rm d}x}{E(x)}\right] \;,
\ee
which for a flat universe ($\Omega_k=0$) reduces to:
\be
d_{L} =  \frac{c(1+z)}{H_0} \int_0^z \frac{{\rm d}x}{E(x)} \;.
\ee
The main result of numerous studies using this procedure is that distant 
SNIa's are on average dimmer by $\sim$0.2 mag than expected 
in an Einstein-deSitter model, which translates in them being $\sim 10\%$ further
away than expected.

The amazing consequence of these results is that 
we live in an accelerating phase
of the expansion of the Universe, an assertion that needs to be
scrutinized on all possible levels, one of which is to verify the
accelerated expansion of the Universe using alternative to SNIa's
extragalactic standard candles. Furthermore, the cause and rate of the
acceleration is of paramount importance, ie., the DE equation
of state is the next fundamental item to search for 
 and to these directions we hope to contribute with our current project.

\subsection{Theoretical Expectations:}
To appreciate the magnitude of the Hubble relation variations 
due to the different DE equation of states, we plot in Figure 1 the relative
deviations of the distance modulus, $\Delta\mu$, of different {\em dark-energy}
models from a nominal {\em standard} (w$=-1$) 
$\Lambda$-cosmology (with $\Omega_m=0.27$ and
$\Omega_{\Lambda}=0.73$), with the relative deviations defined as:
\be
\Delta\mu=\mu_{\Lambda} - \mu_{\rm model} \;.
\ee
The parameters of the different models used are shown in Figure
1. As far as the {\em dark-energy} equation of state parameter is
concerned, we present the deviations from the {\em standard} model 
of two models with a constant w value and of two models with an
evolving equation of state parameter, utilizing the form of eq.(\ref{eqstatez}).
 In the left panel of Figure 1 we present
results for selected values of $\Omega_m$, while in the right panel
we use the same {\em dark-energy} equations of state parameters but
for the same value of $\Omega_m (=0.27)$ (ie., we avoid
the $\Omega_m-{\rm w}(z)$ degeneracy).

Three important observations should be made from Figure 1:
\begin{enumerate}
\item The relative magnitude deviations between the different {\em dark-energy}
  models are quite small (typically $\mincir 0.1$ mag), which puts severe pressure on
  the necessary photometric accuracy of the relevant observations.
\item The largest relative deviations of the distance moduli occur at
redshifts $z\magcir 1.5$, quite larger than those
currently traced by SN Ia samples, and
\item There are strong degeneracies between the different cosmological
  models at redshifts $z\mincir 1$, and in some models even up to
  much higher redshifts (eg., between
  the models with $(\Omega_m, {\rm w}_0, {\rm w}_1)=(0.31,-1,0)$ and 
$(0.29,-1,0.3)$; see Figure 1).
\end{enumerate}
Luckily, as discussed already in the introduction, such degeneracies can be broken
by using other cosmological probes
(eg. the clustering of extragalactic sources, the CMB shift parameter, BAO's, etc). Indeed,
current evidence overwhelmingly show that the total matter content of the universe
is within the range: $0.2\mincir \Omega_m \mincir 0.3$, a fact that reduces significantly
the degeneracies between the cosmological parameters. 
\begin{figure*}
\centering
\mbox{\epsfxsize=16cm \epsffile{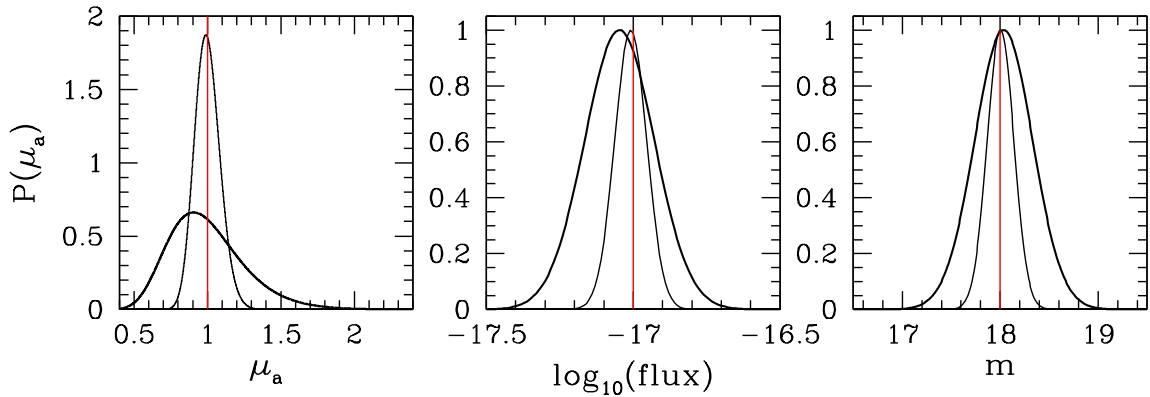}}
\caption{Probability density function of the lensing magnification (left), flux (middle) 
and magnitude (right)
distributions for a source with intrinsic magnitude uncertainty of $\sigma_{\rm int}=0.1$ at 
two different
redshifts (thin line corresponds to $z=1$, while the thick line to $z=3$).}
\end{figure*}

\subsection{Gravitational Lensing Effects on High $z$ Distance Moduli}
A potentially important systematic effect that could hinder attempts to put stringent
cosmological constraints via the Hubble relation, especially when
using high $z$ standard candles 
(which as we saw are those precisely that differentiate between DE equations
of state) is related to gravitational lensing by structures intervening between source and 
receptor.
It is indeed known that the gravitational potential of large-scale structure affects the propagation 
of light from high redshift sources and thus also the distance modulus
of similarly high redshift 
standard candles (eg., Holz \& Wald 1998; Holz \& Linder 2005; Brouzakis \& Tetradis 2008 and 
references therein). These studies assume a Robertson-Walker background superimposing 
a locally inhomogeneous universe and take into account both strong and weak lensing effects. The
resulting magnification distribution of a single source over different paths 
is non-Gaussian and therefore has a non-trivial effect on its distance modulus, especially 
so for the high $z$ standard candles.
 
The main characteristics of the magnification probability density function, $P(\mu_a)$, as derived 
from a variety of studies based on Monte-Carlo analyses and ray-tracing techniques, is that 
$P(\mu_a)$ resembles a log-normal distribution with zero mean (the mean flux  of each source 
over all possible different paths is conserved, since lensing does not affect photon numbers), 
with a mode shifted towards the de-magnified regime and a long tail to high magnification. This 
implies that most sources will be de-magnified, inducing an apparently 
enhanced accelerated expansion, while a few will be highly magnified. The effect is obviously
stronger for higher redshift sources since the lower the redshift the less the optical 
depth of lensing. Note that although the detailed shape of $P(\mu_a)$ is a function of the
underlying cosmology, density profile and evolutionary phase of the intervening
cosmic structures, the main features discussed previously remain unaltered (eg., Wang et al. 2002).

We will therefore model the lensing effect using a log-normal magnification distribution, according 
to appendix A of Holz \& Linder(2005; HL05 hereafter). The fact that the mean flux, over 
all different paths of a source, converges to the unlensed value
implies that if we had a large
 number of standard candles densely populating all the redshift bins, the lensing effects 
would be smoothed out and it would be unnecessary to correct. However, this is not 
usually the case and therefore we need to take lensing into account
(especially for the high $z$ sources). 
\begin{figure*}
\centering
\mbox{\epsfxsize=16cm \epsffile{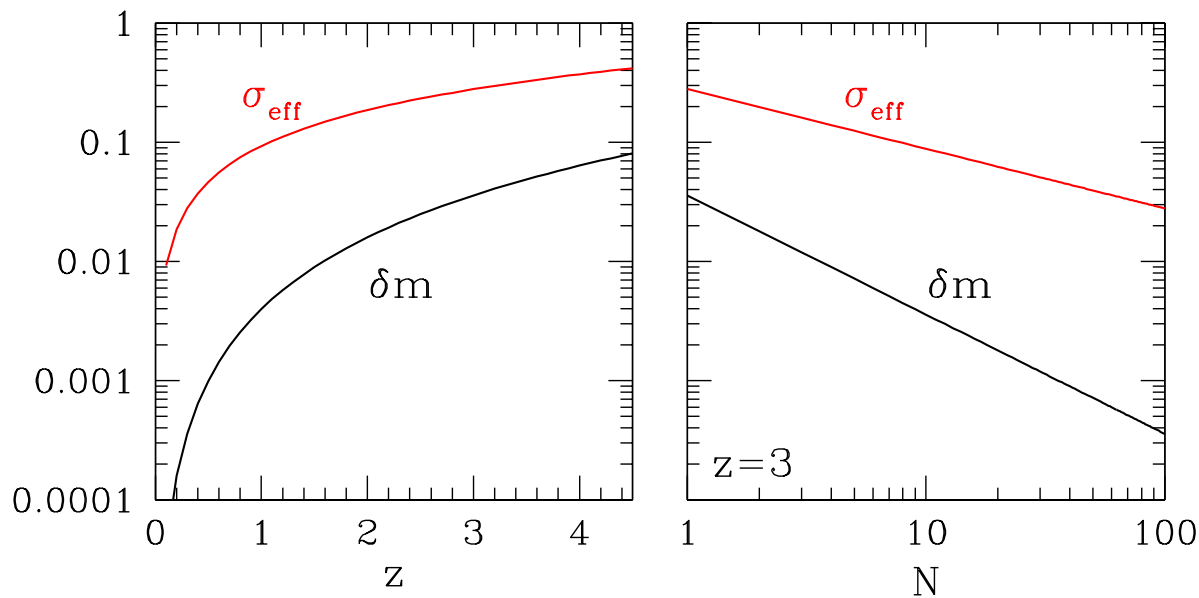}}
\caption{The mode offset ($\delta m$) and effective variance ($\sigma_{\rm eff}$) (a) of a source as
a function of redshift (left panel) and (b) as a function of the number of sources, $N$, 
tracing the same redshift (for this example $z=3$).}
\end{figure*}
 
\noindent
Two main effects of lensing will be accounted for: 
\begin{itemize}
\item the increase of the distance modulus uncertainty by a further term due to 
lensing, $\sigma_{\rm eff}$, which was found by HL05 to be a linear
function of redshift with $\sigma_{\rm eff}=0.093 z$. For a large number, $N$, of
paths (or equivalently of sources) the lensing distribution is approximately Gaussian with 
variance $\sigma_N^2$ and although the lensing distribution of a single path (source) is non-Gaussian,
 we can define the effective variance of a single path (or source) as: 
$\sigma_{\rm eff}^2 = N \sigma_N^2$.
As suggested by HL05 a reasonable $\sigma_{\rm eff}^2/N$ 
contribution to the total distance modulus variance, is given by requiring $N\magcir 10$ 
within $z$-bins of $\sim 0.1 z$ width. Note that this is the only
lensing dependent effect that has been taken into account in some of the SNIa based
analyses (eg., Kowalski et al. 2008; Amanullah et al. 2010).

\item the shift of the mode of the distance modulus distribution to de-magnified values (fainter) 
due to lensing. This is an effect that has not yet been taken into account
in the SNIa based studies.
\end{itemize}
In order to investigate this later effect, and as we have already
pointed out previously, we will use the log-normal approximation to the
magnification distribution due to its nice analytical properties and its resemblance to the
actual magnification distribution (see HL05). If $\mu_a$ is the
source magnification, then its probability distribution is approximated by:
\begin{equation}
P(\mu_a)=\frac{1}{2\pi} \frac{1}{S \mu_a} \exp{\left[-\frac{(\ln \mu_a - \langle \ln \mu_a\rangle)^2}
{2 S^2}\right]}
\label{eq:logn}
\end{equation}
with $S^2$ the variance of $\ln \mu_a$.
The mean magnification is given by $\langle \mu_a\rangle=\exp{(\langle \ln \mu_a\rangle + S^2/2)}=1$,
implying that $\langle \ln \mu_a\rangle =-S^2/2$. Therefore, the probability function is skewed
($\langle \ln \mu_a \rangle < 0$) and determined by only the $S$ parameter. 
From eq.(\ref{eq:logn}) HL05 
derived the corresponding flux distribution, which is also log-normal, and then the 
corresponding magnitude distribution, which is given by:
\begin{equation}
P(m)=\frac{1}{\sigma_m\sqrt{2 \pi}} \exp{\left[\frac{(-m-m_0+b\langle \ln \mu_a\rangle)^2}
{2\sigma_m^2}\right]}
\end{equation}
with $b=2.5/\ln10$ and $\sigma_m^2=\sigma_{\rm obs}^2+(b S)^2$. 
Therefore, we re-capitulate that the effects of lensing are:

\noindent
(a) an offset of the mean, given by:
\begin{equation}
\langle m \rangle = m_0 + \delta m
\end{equation}
where $m_0$ is the intrinsic (de-magnified) magnitude, and
$\delta m = - b\langle \ln \mu_a\rangle=b S^2/2$, 
and, 

\noindent
(b) an increase of the variance 
for which we have that $\sigma^2_{\rm eff}=(b S)^2$ and thus $S^2=\sigma_{\rm eff}^2/b^2$.

Recalling that for large enough sources ($N$), in relatively small $z$-bins ($\sim 0.1z$), 
we have that $\sigma^2_{\rm eff}=N \sigma^2_{N}$, we obtain that the magnitude offset of sources 
within the redshift bin is given by:
\begin{equation}
\delta m (z) = \frac{\sigma_{\rm eff}^2}{2 b N} = \frac{(0.093 z)^2}{2 b N}
\label{eq:offset}
\end{equation}
and the total source variance: 
\begin{equation}
\sigma^2_m(z)=\sigma^2_{\rm obs}+(0.093 z)^2/N \;.
\label{eq:sigma}
\end{equation}

In Figure 2 we plot as an example the magnification, flux and magnitude 
probability distributions for 
the case of a single source placed at two different redshifts ($z=1$ and 3) 
and for an intrinsic observational magnitude uncertainty of $\sigma_{\rm obs}=0.1$.
The intrinsic flux of the source is assumed to be $10^{-17}$ erg s$^{-1}$ cm$^{-2}$
 and the intrinsic 
apparent magnitude $m=18$. The main effects discussed previously are clearly seen, ie., a mode offset 
towards the demagnificaton (fainter) regime and an enhanced variance, both increasing with redshift.
In Figure 3 we plot the expected increase of both the mode offset and 
variance (in magnitudes) as a function of redshift for a single source (left panel). 
In the right panel of the same Figure we plot the suppression of both quantities as we 
increase the source sampling (the case shown corresponds to a source located at a redshift $z=3$).

We will therefore correct statistically the distance moduli of observed standard candles (SNIa, GRBs,
HII-galaxies, etc) by subtracting  an offset $\delta m(z)$ from their raw distance modulus 
(according to eq.\ref{eq:offset}), 
within redshift bins of $\sim 0.1z$ width and using as the
total distance modulus uncertainty that given by eq.(\ref{eq:sigma}).

\subsection{Best Strategy to Determine the DE Equation of State}
\subsubsection{Fitting Models to the Data}
We can now proceed with our investigation to find an efficient strategy to
put more stringent constraints
on the {\em dark-energy} equation of state. To this end we have decided to re-analyse two 
recently compiled
SNIa samples, the Davis et al. (2007) [hereafter {\em D07}] compilation of 192 SNIa
(based on data from Wood-Vasey et al. 2007, Riess et al. 2007 and
Astier et al. 2007) and the
{\em Constitution} compilation of 397 SNIa (Hicken et al. 2009).
Note that the two
samples are not independent since most of the {\em D07} is included in
the {\em Constitution} sample.

Firstly, we present in the left panel of Figure 4 the {\em Constitution} SNIa distance moduli
overploted (red-continuous line) with the theoretical expectation of a flat cosmology
with $(\Omega_m, {\rm w})=(0.27,-1)$. In the inset we
plot the distance moduli difference between the SNIa data and the previously
mentioned model. To appreciate the level of accuracy needed in order
to put constraints on the equation of state parameter, we also plot the
distance moduli difference between the reference $(\Omega_m, {\rm
  w})=(0.27,-1)$ and the $(\Omega_m, {\rm w})=(0.27,-0.85)$ models (continuous red and dashed blue line respectively).

\begin{figure}
\centering
\mbox{\epsfxsize=8cm \epsffile{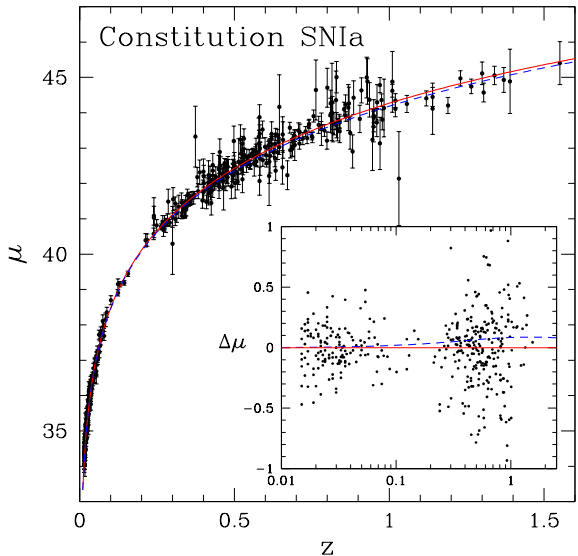}}
\caption{SNIa distance moduli as a function of redshift. {\em
    Inset Panel:} Distance moduli difference between the
best fit model (see Table 1) and the SNIa data. The blue dashed line is the
  corresponding difference between the reference (red-continuous line) ($\Omega_m,$w)$=(0.3,-1.04)$ 
and the ($\Omega_m,$w)$=(0.3,-0.85)$ {\em dark-energy} models.}
\end{figure}

We proceed to analyse the SNIa data by defining the usual likelihood
estimator\footnote{Likelihoods
  are normalized to their maximum values.} as:
\be
{\cal L}^{\rm SNIa}({\bf p})\propto {\rm exp}[-\chi^{2}_{\rm
  SNIa}({\bf p})/2]
\ee
where ${\bf p}$ is a vector containing the cosmological 
parameters that we want to fit for, and
\be
\chi^{2}_{\rm SNIa}({\bf p})=\sum_{i=1}^{N} \left[ \frac{ \mu^{\rm th}(z_{i},{\bf p})-\mu^{\rm obs}(z_{i}) }
{\sigma_{i}} \right]^{2} \;\;,
\ee 
where $\mu^{\rm th}$ is given by eq.(\ref{eq:dm}), $z_{i}$ is the observed
redshift and $\sigma_{i}$ is the 
distance modulus uncertainty, which includes the observational uncertainty and 
the gravitational lensing variance (see eq. \ref{eq:sigma}). 
Since in occasions the observational distance
modulus uncertainty has the form: $\mu^{+\sigma_{p}}_{-\sigma_{n}}$, ie., it
is non-symmetric (due to its logarithmic dependence on the flux), we will use a 
slightly different weighting scheme in the minimization function that takes into account
the asymmetric observational uncertainty.
Following Barlow (2004), and assuming that the likelihood function of the observed $\mu$, derived from
the theoretical $\mu({\bf p})$, is a Gaussian, we can use the following weighting scheme of 
the $\chi^2$ function:
\be
\sigma_i=\sigma_1+\sigma_2 \left[\mu^{\rm th}({\bf p})-\mu^{\rm obs}\right]
\ee
with $\sigma_1=2\sigma_{p}\sigma_{n}/(\sigma_{p}+\sigma_{n})$ and 
$\sigma_2=(\sigma_{p}-\sigma_{n})/(\sigma_{p}+\sigma_{n})$. Obviously, when $\sigma_{p}=\sigma_{n}$ we 
recover the usual symmetric error weighting.
 
In what follows we will constrain our analysis within the framework 
of a flat ($\Omega_{m}+\Omega_{Q}=1$) cosmology and therefore 
${\bf p}\equiv (\Omega_m, {\rm
  w}_0, {\rm w}_1)$. Note that we sample the various parameters on a
grid as follows: the matter density $\Omega_m \in [0.04, 0.64]$, the
equation of state parameter $w \in [-2.0, -0.5]$, while when using a
time-dependent equation of state: $w_0 \in [0, -2]$ and $w_1 \in [-3,
3]$. The typical step size that we use is 0.0015.
Note that the uncertainty of each fitted parameter will be
estimated after marginalizing one parameter over the other, providing as its
  uncertainty the range for which $\Delta \chi^2\le 2.3$
  ($2\sigma$). Such a definition, however, may hide the extent of a
  possible degeneracy
  between the two fitted parameters and therefore it is important to
  visualize the 2D solution space, as indicated in the relevant contour plots.

\begin{figure}
\centering
\mbox{\epsfxsize=8cm \epsffile{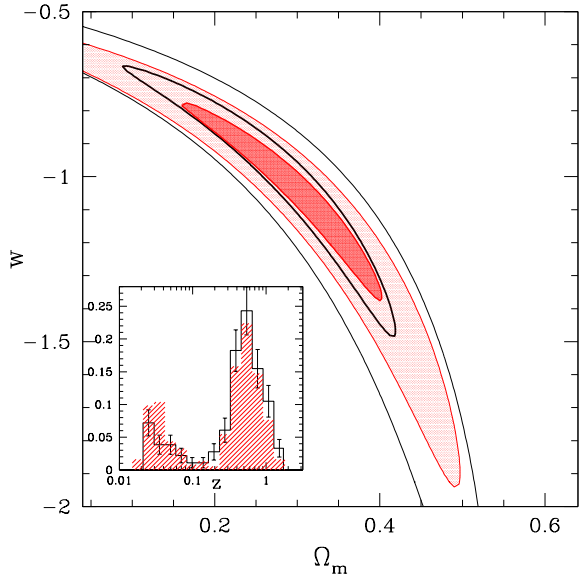}}
\caption{Cosmological parameter solution space 
  using either of the two SNIa data sets ({\em Constitution}: red shaded
  contours and {\em D07}: black contours). Contours corresponding to
  the 1 and 3$\sigma$ confidence levels are
shown (ie., plotted where $-2{\rm ln}{\cal L}/{\cal L}_{\rm max}$ is equal
to 2.30 and 11.83, respectively). 
{\em Inset Panel:} Normalized redshift distributions of the two SNIa data
sets (the shaded histogram corresponds to the {\em Constitution} set).}
\end{figure} 

\begin{figure}
\centering
\mbox{\epsfxsize=8cm \epsffile{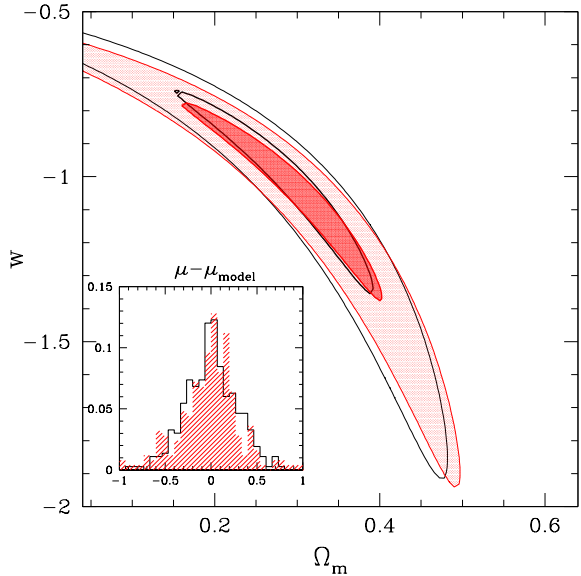}}
\caption{Comparison between the {\em Constitution} SNIa
  constraints (red shaded contours) and those derived by a Monte-Carlo 
procedure designed to closely
  reproduce them (for clarity we show only contours corresponding to 1
  and 3 $\sigma$ confidence levels). {\em Inset Panel:} The {\em
    Constitution} SNIa distance modulus
  deviations from the best fit model $(\Omega_m, {\rm w})\simeq
  (0.30,-1.01)$; see Table 1 -
  and a random realization of the model deviations (red shaded §histogram).}
\end{figure} 

\subsubsection{Larger Numbers?} 
The first issue that we wish to address is how better have we done in
imposing cosmological constraints by increasing the available SNIa
sample from 181 to 366 (excluding the $z<0.02$ SNIa)\footnote{
We use only SNIa with $z\ge 0.02$ in order to avoid
redshift uncertainties due to the local bulk flow (eg. Hudson et
al. 1999 and references therein).}, ie., more than doubling the
sample. 
Table 1 presents various solutions
using each of the two previously mentioned samples. 
Note that since only the relative distances of the SNIa are accurate and not their
absolute local calibration, we always marginalize with respect to the
internally derived Hubble constant (for methods that 
do not need to {\em a priori} marginalize over the
internally estimated Hubble constant, see for example Wei 2008). 

Regarding the fitted parameters uncertainty, we remind the reader
that the definition we use (see above)
cannot clearly reveal the extent of the degeneracy between the two
parameters. A possible measure of such a degeneracy, beyond
inspecting the relevant contour plots, 
is to also present the whole range of the $1\sigma$ contours 
for each parameter. For example, the corresponding
ranges are: $\Omega_m\in [0.11, 0.42]$ and $\Omega_m\in [0.18,
  0.40]$ for the {\em D07} and {\em Constitution} SNIa data sets
  respectively, while the corresponding w ranges are: 
  w$\in [-0.66, -1.48]$ and w$\in [-0.78, -1.38]$. 

Although the
derived cosmological parameters are consistent between the two data
sets, possibly indicating the robustness of the method, the corresponding
goodness of fit (the reduced $\chi^2$) is significantly 
larger in the case of the {\em Constitution} set (1.21 compared to 1.045 of
the {\em D07} set). 
This appears to be the outcome of the different approaches chosen in
order to join the different contributing SNIa sub-sets. According to
Hicken 2009 (private communication) in the case of the {\em D07} the
nearby SNIa were constrained to provide a $\chi^2/{\rm dof}\simeq 1$ by hand,
while no such fine-tuning was imposed on the {\em UNION} set (on which the
{\em Constitution} set is based). A secondary reason could be that 
the latter set includes
distant SNIa which have typically larger distance modulus uncertainties,
with respect to those used in {\em D07}.
Overall, the higher $\chi^2/$dof value of the {\em Constitution} set
should be attributed to a typically lower uncertainty in $\mu$.
As a crude test, we have increased by 20\% the distance modulus
uncertainty of the {\em Constitution} nearby SNIa ($z\mincir 0.4$) 
and indeed we obtain $\chi^2/{\rm dof}\simeq 1.07$, similar to that of {\em D07}.
To also test whether lensing could have a significant effect on the derived cosmological values,
we apply our lensing magnification correction procedure (see section
2.2) to both SNIa compilations and find a very small and insignificant
change of the uncorrected for lensing results 
(see Table 1, last two rows), but interestingly a slightly better reduced $\chi^2$ value.
\begin{table*}
\caption{\small Cosmological parameter fits using the SNIa data for a
  flat prior cosmology. Note that for
  the case where ${\bf p}=(\Omega_m, {\rm w})$, the indicated
  uncertainties are estimated by fixing one parameter at its best
  value and allowing the other to vary, providing as its
  uncertainty the range for which $\Delta \chi^2\le 2.3$.}
\tabcolsep 12pt

\begin{tabular}{|ccc|ccc|} \hline
  \multicolumn{3}{c}{\em D07}  &\multicolumn{3}{c}{\em Constitution} \\ \hline 
         w & $\Omega_{m}$            & $\chi^2_{\rm min}$/dof & w          & $\Omega_{m}$            & $\chi^2_{\rm min}$/dof  \\ \hline
\multicolumn{6}{c}{\em Raw} \\ \hline 
${\bf -1}$               & $0.287\pm 0.020$ & 186.721/180 & ${\bf -1}$        & $0.285^{+0.015}_{-0.014}$ &  439.745/365 \\
$-1.005\pm 0.076$ & $0.289\pm 0.030$ & 186.721/179 & $-1.038\pm 0.053$ & $0.300\pm 0.022$        & 439.703/364 \\ \hline
\multicolumn{6}{c}{\em Lensing corrected} \\ \hline 
${\bf -1}$               & $0.288\pm 0.020$ &  184.775/180 & ${\bf -1}$ & $0.284\pm 0.014$ &  438.263/365 \\
$-0.995\pm 0.075$ & $0.286\pm 0.030$ & 184.775/179 &$-1.036\pm 0.053$  & $0.299\pm 0.022$ & 438.229/364
\\ \hline
\end{tabular}

\end{table*}

\begin{figure*}
\centering
\mbox{\epsfxsize=17cm \epsffile{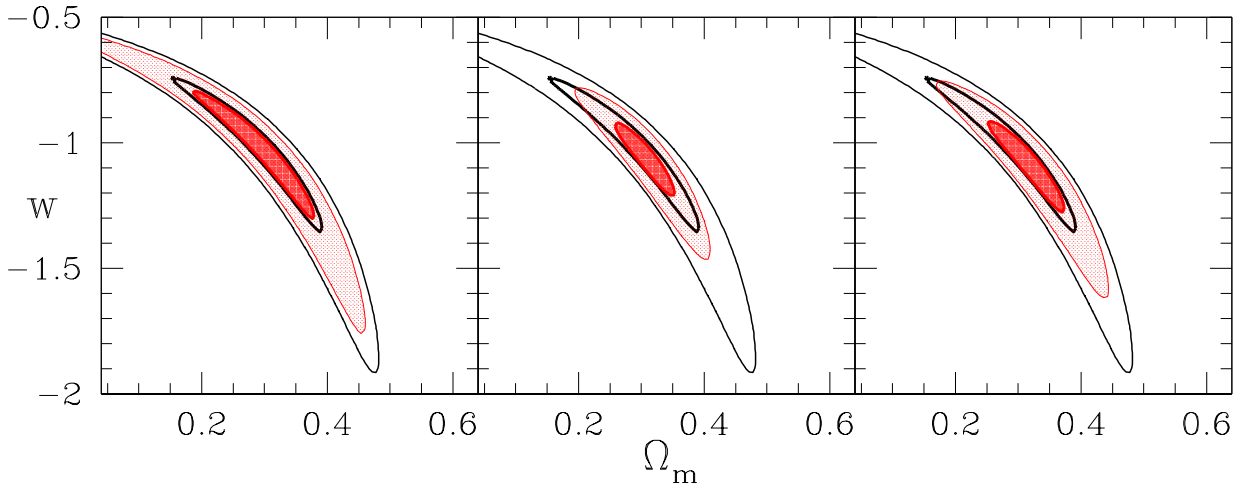}}
\caption{Comparison of the model {\em Constitution} SNIa 
  constraints (black contours) with those (filled contours) derived by reducing to half
  their uncertainties ({\em left panel}), with those derived by adding a sample of
 76 high $z$ tracers ($2\mincir z \mincir 3.5$) with a distance
 modulus mean uncertainty of $\langle \sigma_\mu\rangle \simeq 0.5$
 and no lensing degradation ({\em central panel}), and with those by 
including statistically the expected lensing degradation ({\em right panel}). 
For clarity we show only contours corresponding to the 1  and 3 $\sigma$ confidence levels.}
\end{figure*} 

In Figure 5 we can also see that although the SNIa sample has doubled in 
size, the well-known {\em banana}
shape region of the ($\Omega_m,$w) solution space, indicating the
degeneracy between the two cosmological parameters, is roughly the same for
both data sets. However, there is a reduction of the size of
the solution space when using the {\em Constitution} SNIa
compilation (see also Table 1) at roughly the level expected
from Poisson statistics. 

A first conclusion is therefore that the increase by $\sim 100\%$
  of the Constitution sample has not broken the degeneracy in the
  ($\Omega_m$, w) plane and thus has not provided significantly 
more stringent constraints to the cosmological parameters.
We have further verified that the larger number of SNIa's in the {\em Constitution} sample are
not preferentially located at low-$z$'s (see inset panel of Fig.5) -
in which case we would not have expected more stringent
cosmological constraints using the latter SNIa sample, but they 
have very similar $z$-distributions.

We already have a strong hint, from the previously
presented comparison between the {\em D07} and {\em Constitution}
results, that increasing the number of Hubble
relation tracers, covering the same redshift range and with the current
level of uncertainties, as in the available SNIa samples, 
does not appear to be an efficient avenue for providing stringent
constraints of the cosmological parameters. 

\subsubsection{Lower uncertainties or higher $z$'s:}
We now resort to a Monte-Carlo procedure to
investigate which of the following two directions, that bracket many
different possibilities, provide the required 
more stringent cosmological constraints:

\begin{itemize}
\item Reduce significantly the distance modulus uncertainties of SNIa, 
  tracing however the same redshift range as the currently available samples, or
\item use tracers of the Hubble relation located at 
  redshifts where the models show their largest relative differences
  ($z \magcir 2$), with distance modulus uncertainties comparable to that of
  the highest redshift SNIa's ($\langle \sigma_{\mu}\rangle \simeq 0.4$). At such large 
  redshifts however we expect that the gravitational lensing
  magnification/de-magnification effects will be significant and therefore 
  we will also use the algorithm presented in section 2.2 to statistically degrade
  the intrinsic source flux and investigate its effects on the derived cosmological parameters.
\end{itemize}

The Monte-Carlo procedure is based on using the observed high $z$ SNIa
distance modulus uncertainty 
distribution ($\sigma_\mu$) and a model to assign random $\mu$-deviations from a
reference $H(z)$ function, that reproduces exactly the original
banana-shaped contours of the $(\Omega_m, {\rm w})$ solution space of Figure
5, or in the case of the CPL model of the DE
equation of state the corresponding contours in the $w_0, w_1$
solution-space. Indeed, after a trial and error procedure we have
found that
by assigning to each SNIa (using their true redshift) a distance
modulus deviation ($\delta \mu$) from a reference model
having a Gaussian distribution with zero mean
and variance given by the observed $\langle \sigma_\mu \rangle^2$, 
and using as the relevant individual distance modulus uncertainty
the following: $\sigma^2_i=\sqrt{(1.2
\delta\mu_i)^2 + \phi^2}$ (with $\phi$ a random Poisson
deviate within $[-0.01, 0.01]$) we reproduce exactly the banana-shaped
solution range of the reference model.
This can be seen clearly in Figure 6, where we plot the original {\em
  Constitution} SNIa
solution space (red shaded contours) and the model solution space (black contours).
In the inset panel we show the distribution of the true SNIa
deviations from the best fitted model as well as a random
realization of the corresponding model deviations.

Armed with the above procedure we can now address the questions posed
previously. Firstly, we reduce to half the random deviations of the
SNIa distance moduli from the
reference model (with the corresponding reduction of the relevant
uncertainty, $\sigma_i$). The results of the likelihood analysis can be seen in the
left panel of Figure 7. There is a reduction of the range of
the solution space, but indeed quite a small one. 
Secondly, we add to the {\em Constitution} SNIa sample, a mock
subsample of 76 high $z$ tracers with a
 distance modulus mean uncertainty of $\langle \sigma_\mu\rangle \simeq 0.5$
 (corresponding to that of the current HII-galaxy data) randomly
distributed between $2\mincir z\mincir 3.5$, ie., in a range where the largest
deviations between the different cosmological models occur (see Figure 1).
We now find a significantly reduced solution space (central panel
of Figure 7), which shows that indeed by increasing the $H(z)$
tracers by a few tens, at those redshifts where the largest deviations
between models occur, can have a significant impact on the recovered
cosmological parameter solution space. If we include the expected
lensing degradation of the distance modulus (according to
eq.\ref{eq:offset}), then we observe (right panel of Figure 7) a
slightly worsening of the solution space, but still significantly
smaller than that of the left panel of Fig.7.

The main conclusion of the previous analysis is that a 
more efficient strategy
to decrease the uncertainties of the cosmological parameters, based on
the Hubble relation, is to use standard candles which trace also the redshift
range $2\mincir z \mincir 3.5$. However, in such a case the effects of
gravitational lensing can be severe, especially for small number of
high $z$ tracers, and therefore it is necessary to be taken into account.
\begin{figure}
\centering
\mbox{\epsfxsize=8cm \epsffile{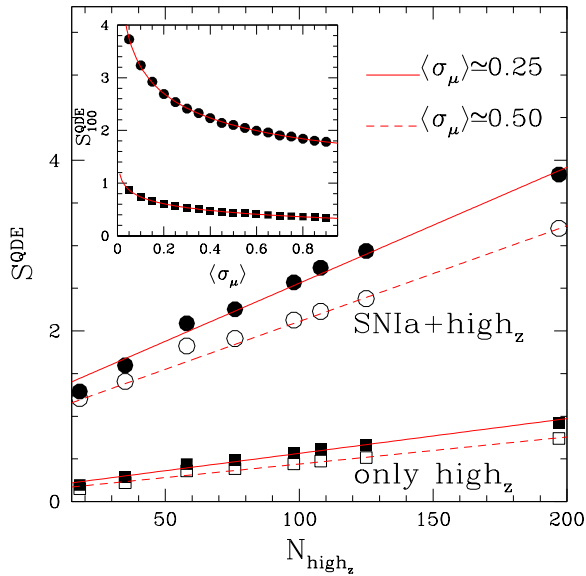}}
\caption{The ``reduction'' parameter $S$, indicating 
the factor by which we reduce the $2\sigma$ contour area of the cosmological
parameters ($\Omega_m,$w) solution space (QDE model) as a function of the number 
of high $z$ tracers ($2<z<3.5$)  of the Hubble relation and for two
different values of the mean intrinsic distance modulus scatter (as
indicated in the plot).
Circular points correspond to using the high $z$ tracers together with 
the current best SNIa data set, while the squares to using
only the high $z$ tracers (and a local $z<0.2$ {\em calibration} sample).
{\em Inset Panel:} The ``reduction'' parameter for the case of using
100 high $z$ tracers as a function of the mean distance modulus
uncertainty, $\langle \sigma_{\mu} \rangle$. The lines correspond to
logarithmic fits to the data (see text).
}
\end{figure} 

\subsubsection{Figure of Merit Analysis}
In order to study the relation between the number of high $z$ tracers
used and the corresponding  reduction of the cosmological parameter
solution space, we will use the Figure of Merit 
(FoM; Bassett 2005; Albrecht et al. 2006; Bueno Sanchez, Nesseris \& 
Perivolaropoulos 2009), defined as the reciprocal area of the 
2$\sigma$ contour (ie., where $-2{\rm ln}{\cal L}/{\cal L}_{\rm max}=6.14$) 
in the parameter space of any two degenerate cosmological parameters
[eg., ($\Omega_m,$ w) for the QDE model or (w$_0$, w$_1$) for the CPL
model], and which has been found to be a useful measure of the
effectiveness of a data set in constraining cosmological parameters. 
A larger figure of merit indicates a greater accuracy in constraining 
the cosmological parameters.

Here we will use a slightly different quantity, which we call
``reduction factor'' and is indicated by $S$, defined as the ratio of the FoM of 
the SNIa+high $z$ Hubble relation solution to that of only the SNIa
(in both cases we use the {\em Constitution} set), in order to study the question of how 
better can we constrain the cosmological parameter space, when adding
$N_{\rm high_z}$ high $z$ tracers of the 
Hubble relation, with respect to the 
best current SNIa data set as a function of the number of high $z$ 
tracers. For example a value $S=2$ 
indicates that the FoM based on the SNIa+high-$z$ Hubble relation is
half of that based on the {\em Constitution}  SNIa data set,
ie., the $2\sigma$ range of the solution space is reduced by a factor of 2.

In Figure 8 we present the results of our analysis
as a function of the number of high-$z$ tracers for
the QDE model.
We present results only for the realistic lensing degradation 
case and for two different values of the distance modulus mean 
observational uncertainty, ranging between a pessimistic 
($\langle \sigma_\mu\rangle=0.5$; open circles)
and an optimistic  ($\langle \sigma_\mu\rangle=0.25$; filled points) value. 

It is evident that including even a small number of high-$z$ tracers 
we can reduce significantly the cosmological 
parameter solution space. 
There is a roughly linear relation between $S$ and the number of 
high-$z$ tracers used, $N_{\rm high_z}$, which depends obviously
on the distance modulus mean uncertainty, 
$\langle \sigma_{\mu} \rangle$ and on whether one uses in addition to
the high-$z$ tracers also the lower redshift SNIa data.

It is also interesting to note that the high-$z$ HII
galaxies could constrain cosmological parameters with the level of
accuracy provided by current SNIa data sets (for $N_{\rm  high_z} \magcir 200$) 
and thus lift any doubts that arise from the fact that SNIa are the only reliable
tracers of the Hubble relation used to-date. Of course the above
relatively large number of high-$z$ HII galaxies can be significantly 
reduced by including an intermediate
population of HII galaxies, ie., tracing similar depths as the current
SNIa samples ($z \mincir 1$). In such a case the expected values of
$S$ will be intermediate between the {\em only high-$z$} and {\em SNIa-high-$z$}
curves of Figure 8. For example, a realistic case of a sample with 80
$0.2<z<1$ and 60 $z\magcir 2$ HII galaxies and a rms distance
modulus uncertainty of $\sim 0.35$ will provide similar constraints as
the current SNIa based analyses ($S\simeq 1$).

In order to quantify the previous results and provide a tool to estimate 
the number of high-$z$ tracers necessary to reduce the current
SNIa solution space by a given factor, taking into account the 
whole parameter space, we first 
normalize the $S$ values by that given for, say,
$N_{\rm high_z}=100$ ($S_{100}$). We then quantify how $S_{100}$ depends 
on $\langle \sigma_{\mu} \rangle$, which is shown in the inset panel of Figure 8.
The continuous curves are logarithmic fits to the data, which 
are given by the following equations:
\begin{equation}\label{eq:16}
S^{\rm QDE}_{100} \simeq  \left\{ \begin{array}{ll}
     1.87 \log_{10}(\langle \sigma_{\mu}\rangle^{-1}+0.74) +1.28 & \mbox{SNIa+high$_z$}\\
     0.44 \log_{10}(\langle \sigma_{\mu}\rangle^{-1}+0.15) +0.30 &
     \mbox{only high$_z$}
                        \end{array}
                  \right. 
\end{equation}
Then, in order to obtain the number of high $z$ tracers, $N_{\rm high_z}$, necessary
to reduce the cosmological solution space (in the QDE model)
by a factor $S$, we fit the normalized value,
$S/S_{100}$ and find:
\begin{equation}\label{eq:17}
N_{\rm high_z} \simeq  \left\{ \begin{array}{ll}
                        187 \; S/S_{100} -88 & \mbox{SNIa+high$_z$}\\
                        139 \; S/S_{100} -39 & \mbox{only high$_z$}
                        \end{array}
                  \right. 
\end{equation}
which has a typical uncertainty of $\sigma_{N}\simeq \pm 5$ for both the
SNIa+high$_z$ and only high$_z$ cases.
The continuous (red) lines in the left panel of Fig.8 are derived from eq.(\ref{eq:16}) and 
eq.(\ref{eq:17}) for $\langle \sigma_{\mu} \rangle=0.25$ and 0.5, and
it is evident that they reproduce extremely well the observed $S$
values (points).
As an example, we can ask how many high $z$ tracers, with
say $\langle \sigma_{\mu} \rangle=0.4$, do we need 
to add to the current SNIa data set in order to
reduce by a factor of 2 ($S=2$) the current SNIa QDE solution space.
Using eq.(\ref{eq:16}) and eq.(\ref{eq:17}) we find $N_{\rm high_z}\simeq 80$,
which drops to $\sim 60$ for $\langle \sigma_{\mu} \rangle=0.25$. It
is therefore interesting to point-out 
that a reduction by a factor of 2 in the distance
modulus uncertainty, of the high $z$ tracers (which is really a non-trivial aim)
can be compensated by a relatively small increase in the number of
high-$z$ tracers.
 
Repeating the previous analysis for the
 case of an evolving DE equation of state 
(CPL; as in eq. 2), and after marginalizing with respect to $\Omega_m$, we also find 
a reduction of the (w$_0$, w$_1$) solution space, when we include the
high-$z$ tracer subsample (Fig.~9),
 but significantly smaller than that of the QDE parametrization. 
We can again estimate what is the 
necessary number of high-$z$ tracers, $N_{\rm high_z}$, having a mean distance modulus 
error of $\langle \sigma_{\mu} \rangle$, in order 
to reduce the cosmological (w$_0$, w$_1$) solution space by a factor $S$.
Again using a parametrization based on the value of $S$ for $N_{\rm
  high_z}=100$, we have that:
\begin{equation}
S^{\rm CPL}_{100} \simeq  \left\{ \begin{array}{ll}
                  0.49 \log_{10}(\langle \sigma_{\mu}\rangle^{-1}+0.65) +1.09 & \mbox{ SNIa+high$_z$}\\
                  0.20 \log_{10}(\langle \sigma_{\mu}\rangle^{-1}+0.69) +0.22 & \mbox{only high$_z$}
                        \end{array}
                  \right.
\end{equation}
and 
\begin{equation}
N_{\rm high_z} \simeq  \left\{ \begin{array}{ll}
                        404 \; S/S_{100} -300 & \mbox{SNIa+high$_z$}\\
                        211 \; S/S_{100} -106 & \mbox{only high$_z$}
                        \end{array}
                  \right. 
\end{equation}
with a typical uncertainty of $\sigma_N\simeq \pm 17$ and $\pm 7$ for the
{\em SNIa+high$-z$} and {\em only high-$z$} cases, respectively.
These results imply than in order to reduce by a factor of 2 ($S=2$) the
current SNIa CPL solution space using high-$z$ tracers with $\langle
\sigma_{\mu} \rangle =0.35$, one needs $N_{\rm high_z}\simeq 300$ and
1200 for the {\em SNIa+high-$z$} and {\em only high-$z$} tracers case, respectively.
The latter value can be significantly reduced if we include an
intermediate redshift ($0.2\mincir z \mincir 1$) HII sample, as
discussed also for the QDE case previously. For example, using a
sample of 80 such intermediate $z$ HII galaxies reduces this number by
a factor of 2.
In any case, the large number of high-$z$ tracers of the Hubble expansion,
needed to effectively constrain the CPL equation of state, renders this task
rather unrealistic. Therefore, 
in order to provide stringent cosmological constraints for the CPL
model (ie., the values of w$_0$ and w$_1$), it would be 
necessary (a) to combine the high-$z$ Hubble relation with that of
current SNIa data, and (b) to join the Hubble relation analysis with other cosmological
tests, like the one that is an integral part of our proposal, ie., the
clustering of X-ray AGN.
Of course other cosmological probes, like BAOs, can and should be used
as well.

We now draw the main conclusions of our Monte-Carlo analysis:
\begin{itemize}
\item Even a small number of high-$z$ ($2\mincir z \mincir 3.5$)
  tracers of the Hubble expansion can reduce significantly the QDE
  model parameter solution space.
\item For the case of the CPL model, in order to reduce the (w$_0$,
  w$_1$) solution space, provided by the current {\em Constitution}
  SNIa set, by the same amount as in the corresponding QDM model, one
  needs three or more as many high-$z$ Hubble expansion tracers.
\item It appears that the effort to reduce significantly the current
  level of random distance modulus scatter of HII galaxies is not as
  important as it is to increase the number of high-$z$ HII galaxies,
  unless one is able to reduce it to $\langle \sigma_{\mu} \rangle
  \mincir 0.1-0.2$ (as can be seen in the inset panel of Fig.8). 
\end{itemize}

\begin{figure}
\centering
\mbox{\epsfxsize=8cm \epsffile{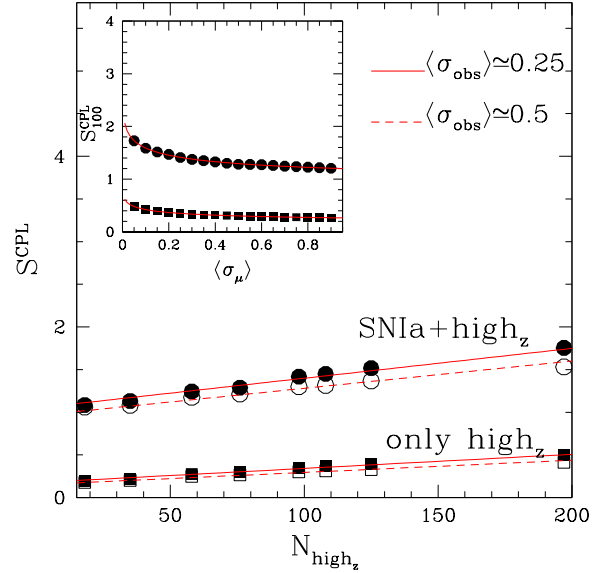}}
\caption{As in Figure 8 but for an evolving DE equation of state (CPL
  model) and after 
marginalizing with respect to $\Omega_m$. The input cosmological model has 
($\Omega_m,$w$_0$, w$_1)=(0.3, -0.98, -0.48)$. The axes scale has been
kept as in Figure 8 in order to appreciate the significant reduction
of efficiency of the high $z$ Hubble relation in providing
cosmological constraints for the CPL model.
}
\end{figure} 

\subsection{A High $z$ Hubble Relation tracer: HII galaxies}
HII galaxies, compact extragalactic objects experiencing massive
bursts of star formation, have a  high luminosity per unit mass, in
large part concentrated in a few strong emission lines in the optical
rest frame. This ensures that the first, obvious, requirement for a
standard candle to be usable at very large distances is met. 

The potential use of HII galaxies as distance indicators stems from
the fact that as one increases the mass of the young stellar
component, not only the ionizing output increases, but also the
turbulent velocity of the gas, which is indicative of supersonic
motions in the gas in the stellar gravitational potential, becomes
larger. This effect induces a correlation between the integrated
luminosity in a nebular hydrogen recombination line, e.g. L(H$\beta$),
which is proportional to the number of ionizing photons, and the line
width $\sigma$. 

Terlevich \& Melnick (1981) found the first
observational confirmation of a correlation between H$\beta$
luminosity and line profile width for giant extragalactic HII
regions and HII galaxies, with residuals that are correlated with
the nebular metallicity.
Subsequent work by Melnick et al. (1987; 1988)
was devoted to obtain a calibration of this correlation in order to
make it suitable for distance measurements. 

The distance indicator, defined as $M_z = \sigma^5 / (O/H)$ with $O/H$
the oxygen abundance of the nebular gas, provides
the predicted luminosity from the relation:
\begin{equation}
\log L(H_{\beta}) = \log M_z + P_0 \;,
\end{equation}
where the zero-point $P_0=29.60$ was originally defined from a sample
of 14 giant extragalactic HII regions (Melnick et al. 1988)
and from which they obtained 
$H_0=80 \pm5$ km sec$^{-1}$ Mpc$^{-1}$. 
Obviously, a critical prerequisite for using such scaling relations as distance 
estimators is an accurate calibration of their zero-points. 
Note that a semi-empirical upper limit of $\sigma=$\,65~km/s has been proposed
for suitable galaxies, which can be explained by the requirement that
HII galaxies are powered by clusters of coeval starbursts with
their dynamics dominated by pressure and not rotation.

The $L(H_{\beta})-\sigma$ relation has been shown to  hold also at 
large redshifts  (Koo et al. 1996, Pettini et al. 2001, Erb et
al. 2003) . Melnick et al. (2000) showed that HII-like starburst
galaxies up to  $z\simeq3$,  satisfy the L(H$\beta$)-$\sigma$
relation, opening the possibility of using the relation to measure
cosmological parameters. They derived the following distance modulus 
relation for HII galaxies:
\begin{equation}\label{eq:ZP}
\mu = 2.5 \log(\sigma^5/F_{H\beta})- 2.5 \log(O/H)- A_{H\beta} +Z_0\;,
\end{equation}
where $F_{H\beta}$ and $A_{H\beta}$ are the flux and extinction in
$H_{\beta}$, respectively. The originally determined zero-point was 
$Z_0=-26.18$ and the rms distance modulus dispersion 
was found to be $\sim 0.52$ mag. 
Although, such an rms uncertainty is larger than what is
obtained with SNIa, the advantage of using HII galaxies is that we
can reach a much larger redshift limit ($z\sim 4$ vs $z\sim 1.7$). 

Using recent galaxy distance
determinations we should be able to better determine the zero-point of
the distance indicator, $Z_0$. To this end we have repeated the original
analysis of Melnick et al. (1988; 2000), using Cepheid and RRLyrae distance
determinations and indeed the rms scatter of the distance
indicator relation is reduced by $\sim 7\%$ while $P_0=29.44$. 
This results in a reduction by $\sim 0.42$ mags of the zero-point (ie., 
$Z_0=-26.60$), which provides results consistent with $H_0=73$ km
sec$^{-1}$ Mpc$^{-1}$ (Ch\'avez et al. 2012 {\em in preparation}). 

It should also be mentioned that there are some  systematic effects 
than can bias distances obtained with the $L(H_{\beta})-\sigma$ 
relation, in particular differences in the ages of the 
stellar populations, contamination from underlying old stellar 
components, or different extinction laws.  To some extent these 
effects can be mitigated by using the equivalent widths of the 
lines to select only very young objects, and the use of modern 
instrumentation that allows a precise control of the size, orientation, 
and location of the spectrograph slits. Still the observations remain 
challenging and require a high level of planning and control.

We are at the process of completing an investigation of these 
effects by using high-resolution spectroscopy of a relatively 
large
number of SDSS low-$z$ HII galaxies with a wide range of relevant 
parameters ($H_{\beta}$ equivalent widths and luminosities, metal 
content, and local
overdensity) in an attempt to understand systematics and to reduce 
the scatter of the distance estimator (Ch\'avez et al 2011 {\em in
  preparation}). 
  
Most high-z HII galaxies known until recently were found in 
broad-band searches aimed mostly to search for Lyman break galaxies, 
which means that they generally have relatively strong continua. Still, a substantial 
fraction present strong emission lines making them 
ideal for our distance estimator (see for example Erb et al. 2006a; 2006b) 
Furthermore, deep slit-less surveys using WFC3 on HST and Narrow band filters at SUBARU have 
revealed substantial numbers of HII galaxies with large 
equivalent widths (i.e. strong emission lines and weak continua) at intermediate and high
 redshifts (Yamada et al. 2005; Kakazu, Cowie, \& Hu 2007; Xia et al.
 2011; Atek et al. 2010; Nestor et al. 2011; Straughn et al.,
 2011). In all, the present sample has more than 400 HII galaxies
 covering the redshift range 0.5$<$z$<$3.7   with about 100 in the
 range 3.0$<$z$<$3.7 and about 150 at  z$\sim $2.

\begin{table}
\caption{\small Cosmological parameter fits using the Siegel et
  al. HII-galaxies and the newly derived zero-point $Z_0$ (eq. \ref{eq:ZP}). 
  The QDE equation of state parameter, w, remains completely
  unconstrained by the current analysis.}
\tabcolsep 16pt
\begin{tabular}{|ccc|} \hline
     w             & $\Omega_{m}$            & $\chi^2_{\rm min}$/dof \\ \hline
${\bf -1}$         & $0.198^{+0.051}_{-0.032}$ & 53.057/14          \\
unconstrained      & $0.280^{+0.048}_{-0.038}$ & 53.849/13  \\ \hline \\
\multicolumn{3}{c}{\em excluding 2 galaxies with tilted emission lines} \\ 
     w             & $\Omega_{m}$            & $\chi^2_{\rm min}$/dof \\ \hline
${\bf -1}$         & $0.224^{+0.063}_{-0.038}$ & 43.119/12          \\
unconstrained      & $0.310^{+0.052}_{-0.046}$ & 42.954/11  \\ \hline \\
\end{tabular}
\end{table}

\begin{figure}
\centering
\mbox{\epsfxsize=8cm \epsffile{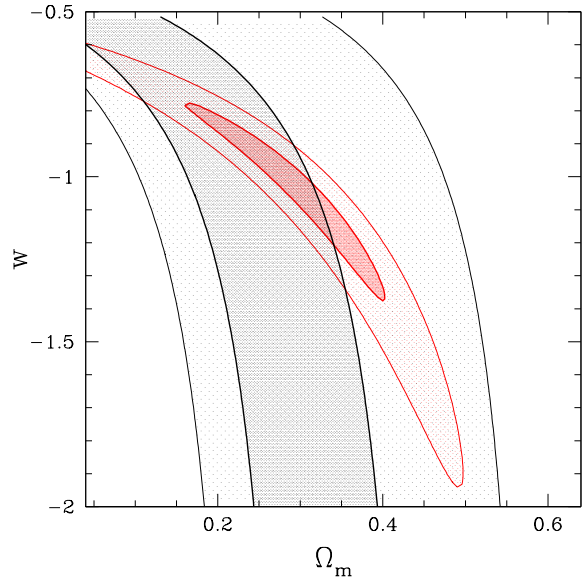}}
\caption{
The HII-galaxy QDE constraints (in the $\Omega_m,$ w plane), 
based on the Siegel et al. sample after excluding two HII galaxies 
showing strong indications for a rotational velocity component.
 Although the constraints are weak, leaving completely
unconstrained the value of w, they are consistent at a
$\sim 1\sigma$ level with the SNIa results (thin red contours).}
\end{figure} 

Summarizing, the use of HII galaxies to trace the Hubble relation, as an
alternative to the traditionally used SN Ia, is based on the following
facts: 
\begin{itemize}
\item[(a)] local and high $z$ HII galaxies define a phenomenological 
relation between $H_{\beta}$ luminosity, velocity dispersion, and 
metallicity as traced by $O/H$ that holds out to cosmological
distances.
 Thus, HII galaxies can be used as alternative tracers of the Hubble 
expansion;

\item[(b)] HII galaxies can be readily observed at much larger
  redshifts 
than those currently probed by SNIa samples, and;
   
\item[(c)] it is at such higher redshifts that the differences between
  the predictions of the different cosmological models appear more 
vividly (see Fig. 1).
\end{itemize}

A more recent application using 15 starburst galaxies with redshift in
the range $z =$\,2.17-3.39 has been carried out by Siegel et
al. (2005)  in an attempt to constrain cosmological
parameters, but the resulting constraints are rather weak. 
They found $\Omega_m =
0.21^{+0.30}_{-0.12}$ for a $\Lambda$-dominated Universe. Clearly, the
errors are still large, and up to this point the results are not
competitive with SNIa. 

We have performed
 our own re-analysis of this data-set, following a similar procedure 
to that applied to the SNIa data in the previous sections, 
but using the newly derived value of $Z_0$ in eq. (\ref{eq:ZP}), allowing 
for the asymmetric uncertainties of the HII-galaxy distance moduli 
and correcting for the effects of gravitational lensing, according 
to section 2.2. Furthermore, we have updated the values of the stellar
velocity dispersion and its uncertainty, for some of the galaxies in
the Siegel et al. sample, according to Erb et al. (2006a).

The resulting constraints on the ($\Omega_m,$w) plane are indicated 
in the first two rows of Table 2, while in the last two rows
 (and in Figure 10)
we present results after excluding two HII galaxies
 (Q1700-MD103 and SSA22a-MD41) that show 
indications of a significant rotational velocity component (derived 
from the tilted emission lines; Erb et al. 2006a), which contaminates 
the estimate of their velocity dispersion. 

Our results show that the derived $\Omega_m$ values, independent of the value 
of w, are towards the lower end of the generally accepted range, while
when excluding the two rotating galaxies the fitted $\Omega_m$ parameter moves towards higher
 values, while there is also a decrease of the value of the corresponding reduced $\chi^2$.

In any case, the main qualitative result of our HII-galaxy based analysis is that 
although the constraints in the $(\Omega_m,$ w) plane are consistent
with those of the {\em Constitution} SNIa analysis, as can be seen in Fig. 10, the
provided cosmological parameter uncertainties are significantly
larger and the degeneracy between $\Omega_m$ and w is even more exacerbated.
These results clearly indicate that the distance indicator for HII
galaxies is highly competitive provided:
\begin{itemize}
\item We increase the number of available high-$z$ ($z\magcir 2$) HII galaxies,
  but it is important to cover also the $0.2<z<1$ range (as shown in section 2.3.4).
\item We apply the estimator to a significant sample of bona-fide high 
redshift HII galaxies selected by the strength of their emission lines 
to ensure no contamination by rotation and/or underlying old  
stellar populations.
\item We minimize all possible sources of systematic and random errors.
\end{itemize}

\section{Cosmological Parameters from the Clustering of X-ray AGN}
The method used to put
cosmological constraints, based on the clustering of some extragalactic
mass-tracer (Matsubara 2004, Basilakos \& Plionis 2009 - hereafter
BP09 - and references therein),
consists in comparing the observed spatial or angular clustering
with that predicted by
different primordial fluctuations power-spectra, in the latter case using
 also Limber's integral equation (Limber 1953) 
to invert the spatial to angular clustering. By minimizing the
differences of the observed and predicted correlation function,
one can constrain the cosmological parameters that enter
in the power-spectrum determination as well as in Limber's inversion.
Using the latter we can relate the
angular and spatial clustering of any extragalactic population under
the assumption of power-law correlations and the small angle
approximation. 

We have chosen X-ray selected AGN as a tracer of the large-scale
structure, in order to perform the previously described analysis, for
the following reasons:
\begin{itemize}
\item[(a)] X-ray selected AGN can be detected out to high redshifts (the peak of their
$z$-distribution is $\sim 1$) and thus trace
the distant density fluctuations providing a further anchor 
of the evolution parameter at a redshift
other than $z\sim 0$, which most galaxy samples trace
to-date. 
\item[(b)] AGN selected through their X-ray emission (and not
in the optical) provide a relatively unbiased census of the
AGN phenomenon, since obscured AGN, largely missed in optical surveys,
are included in X-ray surveys.
\item[(c)] Furthermore, determining the clustering at $\langle z
  \rangle \sim 1$ and $z\sim$0, one can put better 
constraints on the cosmic evolution of the AGN phenomenon and the
evolution of the relation between AGN activity and Dark Matter (DM)
halo hosts (eg. Mo \& White 1996, Sheth et al. 2001), and finally also
on the cosmological parameters and the dark-energy
equation of state (eg. Basilakos \& Plionis 2005; 2006; 2009; 2010).
\end{itemize}

\subsection{Clustering of X-ray AGN: Biases and Systematics}
The earlier ROSAT-based analyses
(eg. Boyle \& Mo 1993; Vikhlinin \& Forman 1995; Carrera et al. 1998;
Akylas, Georgantopoulos \& Plionis, 2000; Mullis et al. 2004)
provided conflicting results on the nature and amplitude of high $z$ AGN
clustering.
With the advent of the XMM and {\em Chandra} X-ray observatories,
many groups have attempted to settle this issue, but in vain.
Different surveys have provided
again a multitude of conflicting results, intensifying the debate
(eg. Yang et al. 2003; Manners et al. 2003;
Basilakos et al. 2004; Gilli et al. 2005; Basilakos et al 2005;
Yang et al. 2006; Puccetti et al. 2006; Miyaji et al. 2007; Gandhi et al.
2006; Carrera et al. 2007; Coil et al. 2009; Starikova et al. 2010). 
However, strong indications exist 
for a flux-limit dependent clustering, interpreted as 
an X-ray luminosity dependent clustering, 
which appears to remove most of the above inconsistencies
(Plionis et al. 2008). Such a luminosity dependent clustering trend
was recently reported also by Cappelluti et al. (2010) and
Krumpe, Miyaji \& Coil (2010).

Furthermore, there  are indications for a quite large high $z$ AGN clustering length,
reaching values $\magcir 10 \;h^{-1}$ Mpc at the brightest flux-limits (eg.,
Basilakos et al 2004; 2005, Puccetti et al. 2006, Plionis et al. 2008;
Cappelluti et al. 2010), which, if verified,
has important consequences for the AGN bias evolution and therefore for the evolution of the
AGN phenomenon (eg. Miyaji et al. 2007; Basilakos, Plionis \&
Ragone-Figueroa 2008 - BPR08 hereafter).
An independent test of these results would be to establish that the
environment of high $z$
AGN is associated with large DM haloes, which being massive should be
more clustered (work in progress).

It is also important to understand and overcome the shortcomings and 
problems that one is facing in
order to reliably and unambiguously determine the clustering properties of the X-ray
selected AGN. Such a list of problems includes the effects of Cosmic
Variance, the so-called amplification bias, the reliability of the $\log
N-\log S$ distribution of the X-ray AGN luminosity function,
etc. (see discussion in Plionis et al. 2009).

Recently, Ebrero et al. (2009a) derived the 
angular correlation function of the soft (0.5-2\,keV) 
X-ray sources using 1063 XMM-{\it Newton} 
observations at high galactic latitudes (2XMM survey).
A full description of the data reduction, source detection and flux 
estimation are presented in Mateos et al. (2008). Note, that the 
survey contains $\sim 30,000$ soft-band point sources within an effective
area of $\sim 125.5$ deg$^{2}$ (for $f_x \ge 1.4 \times
10^{-15}$ erg cm$^{-2}$ s$^{-1}$ ). The large area covered and the
corresponding large number of X-ray sources ensure that the previously
mentioned cosmic variance effects are minimized. However, further
details regarding the
various biases that should be taken into account (the amplification
bias and integral constraint), the survey luminosity
and selection functions as well as issues related to possible non-AGN
contamination, which are estimated to be $\mincir 10\%$, can be found
in Ebrero et al (2009b).

\subsection{Cosmology from the 2XMM angular Clustering}
An optimal approach to unambiguously determine the clustering pattern of X-ray selected AGN
would be to determine both the angular and spatial clustering pattern.
The reason being that various systematic effects or uncertainties enter differently in the
two types of analyses. On the one side, using the angular two-point
correlation function, $w(\theta)$, and its
Limber inversion, one bypasses
the effects of redshift-space distortions and uncertainties related to 
possible misidentification
of the optical counter-parts of X-ray sources. On the other side,
using spectroscopic or accurate photometric redshifts
 to measure the spatial, $\xi(r)$, or projected, $w_p(\theta)$,
 2-point correlation function one by-passes
the inherent necessity, in Limber's inversion of $w(\theta)$, of
assuming a source
redshift-selection function (for the determination of which one uses the
integrated X-ray source
luminosity function, different models of which exist). 

The basic integral equation relating the angular and spatial correlation
functions is:
\begin{equation}
\label{eq:angu}
w(\theta)=2\frac{H_{0}}{c} \int_{0}^{\infty} 
\left(\frac{1}{N}\frac{{\rm d}N}{{\rm d}z} \right)^{2}E(z){\rm d}z 
\int_{0}^{\infty} \xi(r,z) {\rm d}u \;\;,
\end{equation} 
where ${\rm d}N/{\rm d}z$ is the source redshift distribution,
estimated by integrating the appropriate source luminosity function
(in our case that of Ebrero et al. 2009b), folding in also the area
curve of the survey.
Note that to derive the spatial correlation length from eq. (\ref{eq:angu}), 
it is
necessary to model the spatial correlation function as a power
law, assume the small angle approximation as well as a cosmological
background model. The latter is provided by the function $E(z)$
(eq. \ref{eq:HR}), which for a flat background and the QDM equation of state,
it takes the form:
\be 
E(z)=[\Omega_{m}(1+z)^{3}+(1-\Omega_{m})(1+z)^{3(1+{\rm
    w})}]^{1/2}\;.
\ee
The AGN spatial correlation function can be written as:
\be\label{eq:evol}
\xi(r,z) = (1+z)^{-(3+\epsilon)} b^{2}(z) \xi_{\rm DM}(r)\;,
\ee 
where $b(z)$ is the evolution of the linear bias factor (eg. Mo \& White 1997; 
 Matarrese et al. 1997; Sheth \& Tormen 1999; Basilakos \& Plionis
 2001; 2003, Basilakos et al. 2008; Tinker et al. 2010; Ma et al. 2011)
$\epsilon$ is a parameter related to the model
of AGN clustering evolution (eg. de Zotti et al. 1990)
and $\xi_{\rm DM}(r)$ 
is the corresponding correlation function of the underlying dark
matter distribution, given by the Fourier transform of the 
spatial power spectrum $P(k)$ of the matter fluctuations, linearly
extrapolated to the present epoch: 
\be
\xi_{\rm DM}(r)=\frac{1}{2\pi^{2}}
\int_{0}^{\infty} k^{2}P(k) 
\frac{{\rm sin}(kr)}{kr}{\rm d}k \;\;.
\ee
The CDM power spectrum is given by: $P(k)=P_{0} k^{n}T^{2}(\Omega_m,k)$, with
$T(\Omega_m,k)$ the CDM transfer function 
(Bardeen et al. 1986; Sugiyama 1995), $n\simeq 0.967$ and a baryonic
density of $\Omega_{\rm b} h^{2}= 0.02249$, following the 7-year WMAP results (Komatsu et
al. 2011). The
normalization of the power-spectrum, $P_{0}$, can be parametrized by
the rms mass fluctuations on $R_{8}=8 h^{-1}$Mpc scales ($\sigma_8$), 
according to:
\be
P_{0}=2\pi^{2} \sigma_{8}^{2} / \Psi(\Omega_m, R_8)\;,
\ee
with
\be
\Psi(\Omega_m, R_8)= \int_{0}^{\infty}  k^{n+2} T^{2}(\Omega_m,k)
W^{2}(kR_{8}){\rm d}k
\ee
and $W(kR_{8})=3({\rm sin}kR_{8}-kR_{8}{\rm cos}kR_{8})/(kR_{8})^{3}$. 

Evidently, the essential parameters needed to characterize any QDE
cosmological model are: $\Omega_m, w, \sigma_8$ and $H_0$.
Regarding the Hubble constant we will use the WMAP7 results (Komatsu
et al. 2011), which practically coincide with those of 
the HST key project  (Freedman et al. 2001), ie., 
$h=H_0/100=0.704$, while regarding the $\sigma_8$ normalization of the
CDM power spectrum we will use the extrema of the range provided by the recent
analysis of SDSS LRGs for a range of dark
energy equations of state ($\sigma_b \in [0.78,
0.81]$; S\'anchez et al. 2009). Note that the upper limit of the above range
corresponds to the WMAP7 $\Lambda$CDM value (Komatsu et al. 2011).

Furthermore,
to estimate the predicted QDE model correlation function of the underlying mass, 
$\xi_{\rm DM}(r,z)$, in order 
to compare it with the observed AGN clustering, it is necessary to 
deal with the following three issues:
\begin{enumerate}
\item{\bf Clustering Evolution Model:}
As discussed earlier (see eq. \ref{eq:evol}), in order to estimate the
expected clustering of any mass tracer it is important to assume a
clustering evolution model (eg. de Zotti et al. 1990), which is
encapsulated in the value of the parameter $\epsilon$. A value
$\epsilon=-1.2$ corresponds to a 
constant in comoving coordinates clustering model, while a value
$\epsilon=-3$, to a constant in physical coordinates.
According to K\'undic (1997) and Basilakos \& Plionis (2005; 2006) 
we will use the former value of $\epsilon$ 
(although we have also tested the effects of using $\epsilon=-3$).
\item {\bf Bias Evolution Model:} 
We need to calibrate the parameters of the bias evolution model to each
  cosmological model. 
Although a large number of bias evolution models have been proposed in
 the literature (see Papageorgiou, Plionis \& Basilakos 2011, {\em in preparation} for a
 comparison of different models), we use here the approach of 
Basilakos \& Plionis (2001; 2003) which was extended to QDE
cosmological models in BPR08. This model is based on linear perturbation theory and the 
Friedmann-Lemaitre solutions of the cosmological field equations, and 
includes also the effects of interactions and merging of the mass tracers. Its
analytical form has been derived for the QDE cosmological models,
and its generalization to the CPL and alternative gravity
cosmological models is under-way (Basilakos, et al. {\em in preparation}).
Considering that each X-ray AGN is
hosted by a dark matter halo of mass $M_h$, we can analytically predict
its bias evolution behaviour within the QDE models. Conversely fitting
the model to observations, we can determine the mass of the DM halo
within which AGN live (for more details see Basilakos \&
Plionis 2010).

For the case of a spatially flat
cosmological model, our bias evolution model has the following form:
\be
b(M_h,z)={\cal C}_{1}(M_h)E(z)+{\cal C}_{2}(M_h)E(z)I(z)+y_p(z)+1 \;,
\ee
where $y_p(z)$ determines the rate of halo merging\footnote{Note that
the bias factor at the present time is given by: 
$b(M_h, 0)={\cal C}_{1}(M_h)+{\cal C}_{2}(M_h)I(0)+1$.}. However it 
is important only for $z\magcir 3$ and therefore we neglect it here (see BPR08).
Furthermore, we have 
\be 
I(z)=\int_{z}^{\infty} \left(\frac{1+x}{E(x)}\right)^3  {\rm d}x 
\;,
\ee
while the constants $C_1$ and $C_2$ have been fitted using a $\Lambda$CDM simulation
with $\Omega_m=0.3$ and $\sigma_{\Lambda}=0.9$ (BPR08), and have
been found to follow the form:
\be
{\cal C}_{1,2}(M_h)\simeq \alpha_{1,2} \left(\frac{M_h}{10^{13} h^{-1}
M_{\odot}}\right)^{\beta_{1,2}} \;,
\ee
with $\beta_{1}=0.34$, $\beta_{2}=0.32$, while 
$\alpha_1$ and $\alpha_2$ have been found to be cosmological model
dependent, with values given by (Papageorgiou,
Plionis \& Basilakos, 2011 {\em in preparation}):
\be\label{eq:a1}
\alpha_1\simeq \kappa_1 \left(\frac{0.9}{\sigma_{8}}\right)^{\kappa_2} 
\exp{\left[\kappa_3(\Omega_m-0.3)\right]}$$
\ee
with $\kappa_1\simeq 3.44$, $\kappa_2\simeq 2/5$ and $\kappa_3\simeq 4/5$, and
\be
\alpha_2\simeq -0.36 \left(\frac{\Omega_m}{0.3}\right)^{3/2}\;.
\ee

\begin{figure}
\centering
\mbox{\epsfxsize=8cm \epsffile{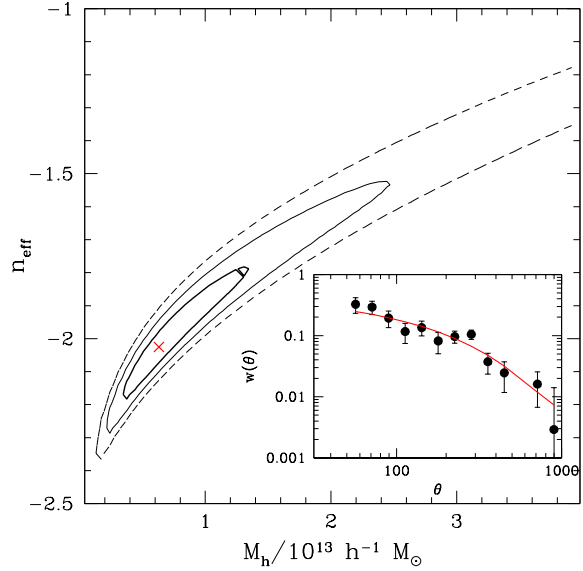}}
\caption{{\em Main Panel:} The 1, 2 and 3$\sigma$ likelihood contours in
  the $M_h, n_{\rm eff}$ parameter space for the $\Lambda$CDM (WMAP7)
  model. {\em Inner Panel:} The Ebrero et al. (2009a) 2XMM angular
correlation function and the best fit $\Lambda$CDM model (continuous line).}
\end{figure}

\item {\bf Non-linear Power Spectrum:}
Since the correlation function on small angular scales is within the
expected non-linear regime,  we should include in our model
power-spectrum the non-linear contributions. To this end we use
the corresponding fitting formula 
introduced by Peacock \& Dodds (1996), for the $\Lambda$CDM
model (see also Smith et al. 2003; Widrow et al. 2009). 
There is one relatively free parameter in their formulation,
which is the slope of the power spectrum at the relevant scales, since
the CDM power-spectrum curves slowly  and thus it varies as a function
of scale: $n_{\rm eff} = {\rm d} \ln P(k)/{\rm d} \ln k$. On the
scales of interest the value is $n_{\rm eff}\simeq -2$, but we have
decided to actually derive, and then fix, the $n_{\rm eff}$ value from
the data analysis itself. Using the minimization procedure discussed
in section 3.3 we compare the observed 2XMM AGN
correlation function with that provided by the WMAP7 $\Lambda$CDM
model (ie., fixing $\Omega_m=0.272$, w$=-1$, and $\sigma_8=0.811$) 
fitting the remaining two free parameters (ie., $M_h$ and $n_{\rm
  eff}$). The corresponding solution space can be seen in Fig. 12,
while the best fit parameter values are: 
$M_h \simeq 6.5 (\pm 2.1) \times 10^{12} h^{-1} M_{\odot}$ and 
$n_{\rm eff}\simeq -2.02^{+0.05}_{-0.04}$. In the inset of Fig.11 we also plot
the 2XMM angular correlation function together with the best fit
$\Lambda$CDM model (continuous line). 

As a consistency check we have
verified that when fixing the non-linear slope of $P(k)$ to the above 
fitted value and leave as
free parameters $M_h$ and $\sigma_8$, we recover the WMPA7
$\sigma_8$ value and exactly the same $M_h$ value, as above. The
derived value of $M_h$ is slightly larger than that provided by Ebrero et al. (2009a)
using the Sheth et al. (2001) bias evolution model (ie., $\simeq 5
\times 10^{12} h^{-1} M_{\odot}$). We have tested also
the case of a clustering evolution model with $\epsilon=-3$, in which
case the derived value of the halo mass is $\simeq 2 \times 10^{10}
h^{-1} M_{\odot}$, a value significantly below any reasonable value that has
been proposed or derived in the literature. We will therefore use
$\epsilon=-1.2$ throughout the rest of the paper.
\end{enumerate}

\subsection{Fitting Models to the 2XMM Clustering Data}
In order to constrain the cosmological parameters 
we use again the standard $\chi^{2}$ 
likelihood procedure and compare the measured 
XMM soft-band source angular correlation function (Ebrero et al. 2009a)
with the predictions of different spatially flat cosmological models.
The corresponding likelihood estimator is defined as:
${\cal L}_{\rm AGN}({\bf p})\propto {\rm exp}[-\chi^{2}_{\rm AGN}({\bf p})/2]$
with:
\be
\label{eq:likel}
\chi^{2}_{\rm AGN}({\bf p})=\sum_{i=1}^{n} \frac{\left[ w_{\rm th}
(\theta_{i},{\bf p})-w_{\rm obs}(\theta_{i}) \right]^{2}}
{\sigma^{2}_{i}+\sigma^{2}_{\theta_{i}}}  \;\;,
\ee 
where ${\bf p} \equiv (\Omega_{m}, {\rm w}, M_h)$, $\sigma_{i}$ is the uncertainty
of the observed angular correlation function and $\sigma_{\theta_{i}}$ corresponds to the width
of the angular separation bins.

We sample the various parameters as follows:
the matter density $\Omega_{m} \in [0.1,0.4]$ in steps of
0.002; the equation of state parameter w$\in [-1.4,-0.6]$ in steps
of 0.005 and the parent dark matter halo
$M_h/10^{13}h^{-1}M_{\odot} \in [0.1,3]$ in steps of 0.01.
In this likelihood analysis we use as priors a flat universe, and the
previously mentioned values of $h$, $\sigma_8$ and $\Omega_{b}$.

The results of the minimization procedure for the case of
$\sigma_8=0.81$ are:
$\Omega_m=0.301\pm 0.008$, w$=-0.990\pm 0.058$ and $M_h=3.1 (\pm 1.1) \times
10^{12}\; h^{-1} \; M_{\odot}$, with a $\chi^2=39.41$ for 10 degrees
of freedom. The large value of $\chi^2/$dof is due to the sinusoidal
modulation of the 2XMM $w(\theta)$, which could be due to systematic
effects possibly related to the size of the XMM fields (see discussion
in BP09). If we use the 2$\sigma$ $w(\theta)$ uncertainty in the
denominator of the $\chi^2$ function of eq. (\ref{eq:likel}), then the
$\chi^2$ drops to $\simeq 9.86$ (and the uncertainties of the fitted
parameters increase to roughly twice the values indicated previously).

These results slightly differ with
respect to the similar analysis of BP09, due to a number of
improvements that we have currently included, apart from the fact that
we have also used the WMAP7 cosmological parameters (ie., $\Omega_b$,
$n$ and $h$). The two main improvements
have to do with (a) the bias evolution model, in which we have now taken into
account the dependence of the parameter $a_1$ (see eq.\ref{eq:a1}) on $\Omega_m$
and $\sigma_8$ (which has mostly affected the derived value of $M_h$,
reducing it significantly), and (b) the non-linear power spectrum
corrections, for which we have used the Peacock \& Dodds (1996)
$\Lambda$CDM fitting formula.

Nevertheless our current procedure can and will be improved in the future in a number of ways: 
\begin{itemize}
\item We will eventually use the
clustering of X-ray selected AGN from a large contiguous X-ray survey,
a fact which will solve, we believe, the quasi-sinusoidal small
amplitude modulation of the 2XMM $w(\theta)$ (see discussion in
BP09). Such a future survey is the XMM-Newton Very Large Programme (XXL), 
which was recently granted time to map two extragalactic regions of 25 deg$^2$, at a
depth of $\sim 5 \times 10^{-15}$ erg cm$^{-2}$ s$^{-1}$ (Pierre et
al. 2011).
\item We will investigate more accurate non-linear power-spectrum
  corrections of $w(\theta)$ (eg., Widrow et al. 2009), and
\item We will ultimately test a large range of Dark Energy
  equations of state, to
  include CPL and alternative gravity [$f(R)$] models. To this
  end we will use the recent generalization of the BPR08 bias
  evolution model of Basilakos, Plionis \& Pouri (2011).
\end{itemize}

\begin{figure*}
\centering
\mbox{\epsfxsize=16cm \epsffile{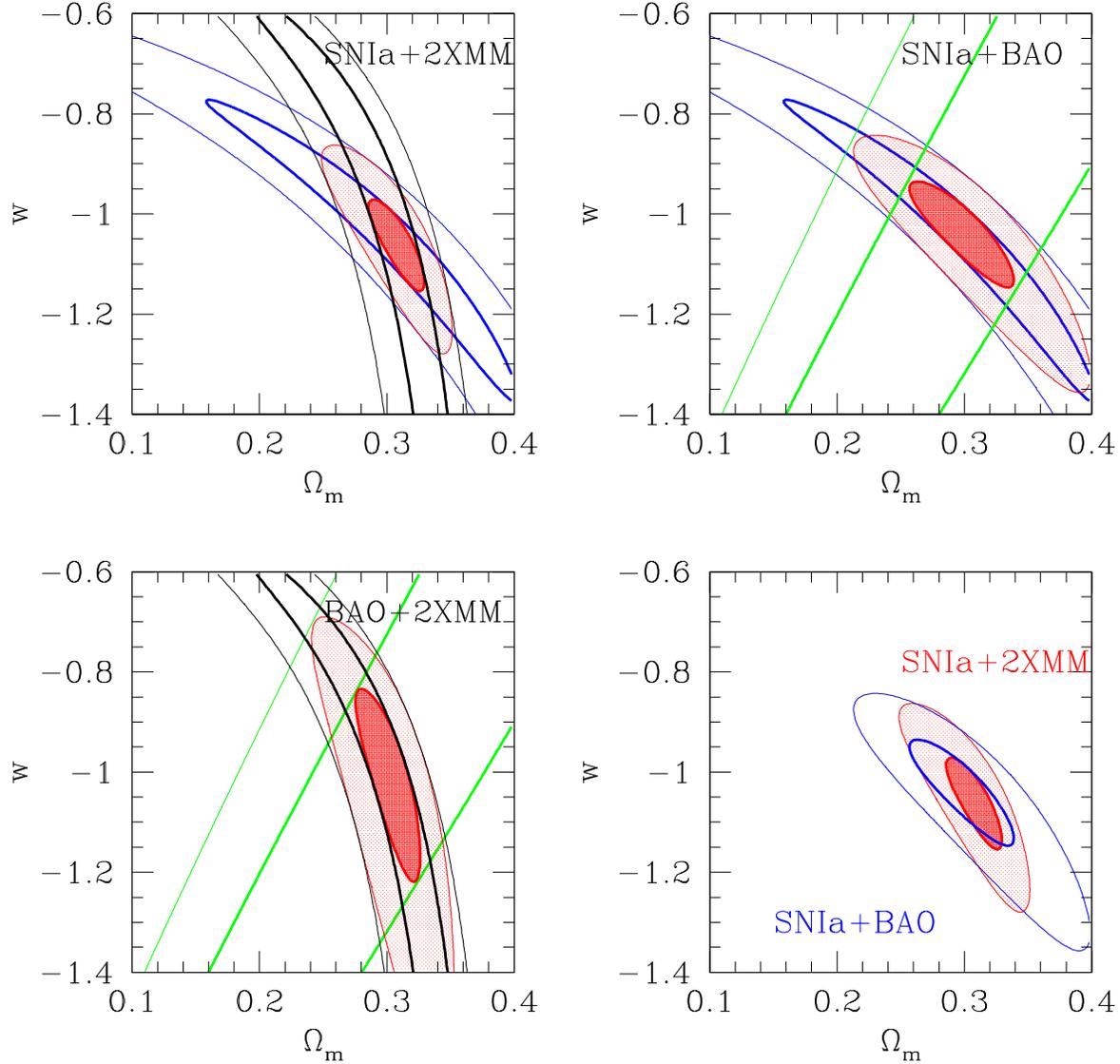}}
\caption{QDE model parameter constraints (ie., in the $\Omega_m,$w
  plane) provided by the joint
  likelihood analysis of pairs of cosmological data. The red shaded
  contours are the joint-likelihood contours of the indicated pairs of
  data. For clarity we show only contours corresponding to the 1
  and 3 $\sigma$ confidence levels. The 2XMM results shown
  correspond to those based on a power spectrum normalization of
  $\sigma_8=0.81$. Note also that we have followed the conservative
approach of using 2$\sigma$ $w(\theta)$ uncertainties in the
minimization process, to allow for
the small amplitude sinusoidal $w(\theta)$ modulation (see main text
and BP09). 
{\em Upper Left Panel:} {\em Constitution} SNIa Hubble relation (blue contours) and
2XMM AGN clustering (black contours). 
{\em Upper Right Panel:} {\em Constitution} SNIa Hubble relation (blue
contours) and LRGs BAO (green contours).
{\em Lower Left Panel:} 2XMM AGN clustering (black contours)
and LRGs BAO (green contours).
{\em Lower Right Panel:} The joint likelihood contours of the
SNIa-2XMM (red contours) and of the SNIa-BAO (blue contours)
pairs.}
\end{figure*}

\begin{table*}
\caption{The best fit values of the cosmological parameters based on
the joint likelihood analysis of the indicated cosmological data. 
For the case of the 2XMM clustering analysis we followed the conservative
approach of using 2$\sigma$ $w(\theta)$ uncertainties.
The uncertainty of each fitted cosmological parameter 
has been estimated after marginalizing over the other parameter (ie.,
 by fixing one parameter at its best value and allowing the other to vary, providing as its
  uncertainty the range for which $\Delta \chi^2\le 2.3$). The
last column indicates the {\em reduction parameter} (related to the
Figure of Merit) as defined in
section 2.3.4.}
\tabcolsep 16pt
\begin{tabular}{|lccrrr|} 
\hline
Joint data   & $\Omega_{\rm m}$          & w &$\chi^2$ & dof & $S$\\ \hline 
SNIa+BAO         &$0.296\pm 0.021$  &$-1.027\pm 0.053$ & 439.711 & 365
& 2.32 \\ \hline

XMM$^a$+SNIa     &$0.310\pm 0.012$  & $-1.064\pm^{0.053}_{0.048}$ & 449.591 & 374 & 4.49\\
XMM$^a$+BAO      &$0.302\pm 0.014$  & $-0.995\pm^{0.096}_{0.128}$ & 9.860 & 11 & 1.74\\
\hline

XMM$^b$+SNIa     &$0.318\pm 0.013$  & $-1.085\pm^{0.059}_{0.048}$ & 449.644 & 374 & 4.35\\
XMM$^b$+BAO      &$0.306\pm 0.015$  & $-0.973\pm^{0.096}_{0.123}$ & 9.861 & 11 & 1.70\\
\hline
\end{tabular}
\\
$^a$ 2XMM results based on a $P(k)$ normalization of $\sigma_8=0.811$,\\
$^b$ based on $\sigma_8=0.78$.
\end{table*}

\section{Joint Hubble-relation, AGN Clustering and BAO analysis}
Here we will perform an example of the joint analysis between the previously discussed  results
from 2XMM clustering and the Hubble relation, which is
the basic aim of our overall project. 
For the current exercise we will use the 
{\em Constitution} SNIa Hubble relation since we are still working
on the development of the HII-galaxy based methodology.
For completeness we will also use
the recent results of the baryonic acoustic oscillation technique (BAO).
We remind the reader that BAOs are produced by pressure (acoustic) waves in the
  photon-baryon plasma in the early universe generated by dark matter
  overdensities. At the recombination era ($z\sim 1100$),
  photons decouple from baryons and free stream while the pressure
  wave stalls. Its frozen scale, which constitutes a standard ruler,
  is equal to the sound horizon length, $r_s\sim 100\; h^{-1}$ Mpc
  (e.g. Eisenstein \& Hu 1998). This appears as a small,
  $\sim 10\%$ excess in the galaxy, cluster, or AGN power spectrum (and
  its Fourier transform, the 2-point correlation
function) at a scale corresponding to $r_s$. First evidence of such an excess
have been reported in the clustering of the SDSS luminous red galaxies (LRGs)
(see Eisenstein et al. 2005, Padmanabhan et al. 2007; 
Percival et al. 2010). 
In this work we use the latest
measurement of Percival et al. (2010):
\[ r_{s}(z_{d})/D_{V}(z_{\star})=0.1390\pm 0.0037 \;,\] 
(see also Kazin et al. 2010a, 2010b). Note that 
$r_{s}(z_d)$ is the comoving sound horizon
size at the baryon drag epoch $z_{d}$, which is given by the 
fitting formula of Eisenstein \& Hu (1998), 
$D_{V}(z_\star)$ is the effective distance measure
and $z_{\star}=0.275$. Of course, the 
quantities $r_{s}, D_{V}$ can be defined analytically, and are given by:
\begin{equation}
r_{s}(z_{d})=\frac{c}{\sqrt{3}}\int_{0}^{a_{d}} 
\frac{da}{a^{2} H(a) \sqrt{1+(3\Omega_{b}/4\Omega_{\gamma})a } }
\end{equation}
where $a_{d}=(1+z_{d})^{-1}$ and $\Omega_{\gamma} h^{2}\simeq 2.47
\times 10^{-5}$ the energy density of photons. 
In this context, the effective distance is (Eisenstein et al. 2005):
\begin{equation}
D_{V}(z)\equiv \left[ (1+z)^{2} D_{A}^{2}(z) \frac{cz}{H(z)}\right]^{1/3}
\end{equation} 
where $D_{A}(z)$ 
is the angular diameter distance. Therefore, the corresponding
$\chi^{2}_{\rm BAO}$ function is simply written as:
\begin{equation}
\chi^{2}_{\rm BAO}({\bf p})=
\frac{[\frac{r_{s}(z_{d})}{D_{V}(z_{\star})}({\bf
p})-0.1390]^{2}}{0.0037^{2}} \;,
\end{equation}
where ${\bf p}$ is the vector containing the cosmological 
parameters that we want to fit for. In this case ${\bf
  p}=(\Omega_m,$w).

We therefore perform a joint likelihood analysis, assuming that any
two pairs of cosmological data sets are independent (which indeed they are) and thus the
joint likelihood can be written as the product of the two individual ones.
The results based on the joint analysis of the different pairs of cosmological
data are shown in Fig.12 and quantified in Table 3. It is evident
that the addition of the XMM clustering analysis provides significantly more stringent
constraints than, for example, the joint SNIa and BAO results. The 
{\em reduction parameter}, ie., the ratio of the Figure
of Merit of the joint XMM-SNIa analysis to that of the {\em Constitution} SNIa analysis (see
definition in section 2.3.4) shows that the $2\sigma$ range of the
$(\Omega_m,$w) solution space is reduced by a factor of $\sim$5/2 
with respect to that of the BAO-SNIa analysis (see lower right panel
of Fig.12).

The necessity, however, 
to impose constraints on a more general, time-evolving, {\em dark-energy} equation 
of state (eq. 2), implies that there is ample space for improving the
current analysis and indeed our aim is to develop further this project
by (a) using a
new Hubble relation analysis, based on high $z$ HII galaxies, as detailed in this
paper, and (b) by generalizing the BPR08 bias
evolution model for any DE equation of state (CPL and alternative
gravity models). 

\section{Conclusion}
We have investigated the question of which is the most
efficient strategy to tighten the cosmological constraints provided
by fitting the Hubble relation. Using extensive Monte-Carlo
simulations we have verified that by using only a few high $z$ tracers
(in the range $2\mincir z \mincir 3.5$), even with a relatively large
distance modulus uncertainty, we can reduce significantly the present
cosmological parameter solution space. 
We have taken into account the effects of
lensing magnification/de-magnification, which not only increases the
distance modulus uncertainty but it also shifts systematically
the mean distance modulus of individual sources. Although the effects
can be severe for an individual source, they can be statistically
treated and they are significantly reduced the denser the source sampling is in redshift space.
Applying our lensing magnification correction to the 
{\em Constitution} SNIa set (Hicken et al.), we find that the
fitted cosmological parameters are not significantlly affected by
such effects, due to the fact that the SNIa sample traces relatively small
redshifts ($z\mincir 1$). 

Based on a figure of merit analysis 
we have provided a simple procedure to estimate the necessary number of
$2\mincir z\mincir 3.5$ tracers needed to reduce the cosmological solution space,
presently provided by the {\em Constitution} set, by a desired factor
of our choice and for any level of rms distance modulus uncertainty.
This analysis has shown that in order to significantly reduce the 
cosmological parameter solution space, it is more efficient to
increase the number of high-$z$ tracers than to reduce their
individual uncertainties.
 A re-analysis of the cosmological constraints provided by a small
sample of high-$z$ ($2.2\mincir z \mincir 3.4$) HII galaxies,
previously analysed by Siegel et al. (2005), but now using a novel
determination of the zero-point of the relevant distance scaling
relation, provides consistent cosmological results with those of SNIa,
although the dark energy EoS parameter remains unconstrained at present.

Finally, using the clustering of X-ray selected AGN 
we provide the
framework that will be used, joining their cosmological likelihood with that of the
Hubble relation analysis, to put stringent dark energy equation
of state constraints.
An example of such a joint analysis, using the
2XMM clustering and the {\em Constitution} SNIa Hubble relation, 
and under the priors of a flat universe, $h=0.704$ and $\sigma_8=0.81$
or 0.78, provide 
significantly more stringent QDE model constraints, as indicated by the
fact that the Figure of Merit increases by a factor $\sim 2$, 
with respect to that of the joint SNIa-BAO analysis.
The QDE cosmological parameters provided by the 2XMM-SNIa joint analysis are:
$\Omega_{\rm m}= 0.31\pm 0.01$ and w$=-1.06\pm 0.05$, with the
uncertainties being estimated after marginalizing one parameter over
the other. 

\subsection*{Acknowledgements}
We thank Dr J. Ebrero for comments and for
providing us with an electronic version of the clustering results
and their XMM survey area curve. 
MP, RT and ET also acknowledges financial
support under Mexican government CONACyT grants 2005-49878, 2005-49847
and 2007-84746 and 2008-103365 respectively.

\end{document}